\documentclass[preprint,showpacs]{revtex4}
\usepackage{epsfig}

\begin{document}

\title{Systematic study of pentaquark states: $qqq-q\bar{q}$ configuration}

\author{Jialun Ping, Hongxia Huang, Chengrong Deng}

\address{Physics Department, Nanjing Normal University, Nanjing, 210097
P.R. China}

\author{Fan Wang}

\address{Center for Theoretical Physics, Nanjing University, Nanjing 210093,
P.R. China}

\author{T. Goldman}

\address{Theoretical Division, LANL, Los Alamos, NM 87545, USA}

\begin{abstract}
Group theoretic method for the systematic study of five-quark states
with meson-baryon ($q\bar{q}-q^3$) configuration is developed. The
calculation of matrix elements of many body Hamiltonian is
simplified by transforming the physical bases (meson-baryon quark
cluster bases) to symmetry bases (group chain classified bases),
where the fractional parentage expansion method can be used. Three
quark models, the naive Glashow-Isgur model, Salamanca chiral quark
model and quark delocalization color screening model, are used to
show the general applicability of the method and general results of
constituent quark models for five-quark states are given. The method
can also be useful in the calculation of meson-baryon scattering and
the study of the five-quark components effect in baryon structure.
The physical contents of different model configurations for the same
multi-quark system can also be compared through the transformation
between different physical bases to the same set of symmetry bases.
\end{abstract}


\maketitle

\section{Introduction}

After 40 years of quark model study, the idea about baryon and meson
is about to go beyond the naive picture: baryon $q^3$ and meson
$q\bar{q}$. The proton spin puzzle could be explained by introducing
the $q^3q\bar{q}$ component in the quark model \cite{qcw}. Recently
to understand the strange magnetic moment of proton, the five-quark
component of proton was proposed \cite{BSZou}. The recent progresses
on meson spectroscopy called for the tetra-quark components
\cite{tetraquark}. Although the pentaquark states claimed by
experimental groups few years ago might be disappeared and the
multi-quark states might be hard to be identified, the multi-quark
study is indispensable for understanding the low energy quantum
chromodynamics (QCD), because the multi-quark states can provide
information unavailable for $q\bar{q}$ meson and $q^3$ baryon,
especially the property of hidden color structure. Generally a
multi-body interaction multi-channel coupling calculation is needed
in the multi-quark study. Therefore a powerful method is necessary.
The fractional parentage (fp) expansion technique has been proven as
a powerful method for few-body problem.

To full play the powerful group theoretic method for quark model
calculation of multi-quark system, one needs not only the fractional
parentage expansion coefficients of these multi-quark states but
also a relation between various quark model states (hereafter called
physical bases) and the group theoretical classification states
(hereafter called symmetry bases). Such a method has been developed
and successfully applied in systematic search of dibaryon
\cite{dibaryon} and five-quark states with Jaffe-Wilczek di-quark
configuration \cite{group1}, where the physical bases were
transformed to symmetry bases first, then the many-body matrix
elements calculation of Hamiltonian (with two body interaction) on
the symmetry bases was done by means of fractional parentage
expansion, {\em i.e.}, the many-body matrix elements can be reduced
to overlap and two-body matrix elements. At last the matrix elements
on the symmetry bases were transformed back to physical bases.

The present work is to provide the transformation coefficients
between physical bases and symmetry bases of five-quark systems to
facilitate the calculation of many body Hamiltonian matrix elements.
The physical bases discussed in this paper is the the meson-baryon
bases. For Jaffe-Wilczek (JW) di-quark bases, see the first paper of
this series \cite{group1}. The method is applied to the quark model
calculation of five-quark systems with three quark models: the naive
quark model, {\em i.e.}, the Glashow-Isgur model \cite{Isgur}, the
Salamanca version of chiral quark model \cite{chiral} and the quark
delocalization color screening model (QDCSM) \cite{dibaryon,qdcsm}.
The general applicability of this method is verified through these
model calculations.

Even though the multi-quark system might be hard to be identified,
the method is still useful in the study of hadron-hadron scattering
and the multi-quark components effect in hadron in the framework of
quark model. The transformation between physical bases and symmetry
bases is also useful in the study of the physical contents of
different model approaches for the same multi-quark system. This
report gave an example where the different model approaches in the
study of pentaquark states are compared.

In section II, the physical bases and symmetry bases are introduced
and the transformation between them is derived. The fractional
parentage technique applied to calculate the matrix elements on the
symmetry bases is also explained in this section. Section III
explains three quark models we used. The results of the systematic
calculation of five-quark system in the $u,d,s$ 3-flavor world are
given in section IV. The last section gives the summary.

\section{Physical bases and symmetry bases}

The physical bases are constructed as follows, first the wave
function of each quark cluster is constructed based on the group
chain classification
\begin{eqnarray}
& & ~[1^n]~~~~~~[\nu]~~~~~~~ [{\tilde
\nu}]~~~~~~~[c]~~~~~~~[\mu]~~~~~~~
[f] ~~~~~~~I~~~~~~Y~~~~~~~~J~~ \nonumber \\
& & \mbox{SU}_{36}\supset \mbox{SU}^x_2\times \left\{
\mbox{SU}_{18}\supset \mbox{SU}^c_3\times \left[ \mbox{SU}_6\supset
\left( \mbox{SU}^f_3\supset \mbox{SU}^{\tau}_2\times \mbox{U}^Y_1
\right) \times \mbox{SU}^{\sigma}_2 \right] \right\}
\label{group-chain}
\end{eqnarray}
(the Young diagrams or quantum numbers for each group are also shown
in Eq.(\ref{group-chain})), then the quark cluster wave functions of
the system was coupled to definite color, spin and isospin quantum
numbers by Clebsch-Gordan (CG) coefficients of color SU$^c_3$, spin
SU$^{\sigma}_2$ and isospin SU$^{\tau}_2$ group, and finally
anti-symmetrized. For meson-baryon ($q\bar{q}-q^3$) configuration,
the five quarks are separated into two clusters: the meson and
baryon clusters with separation $S$. The physical basis
(meson-baryon basis) is defined as
\begin{equation}
\Psi_{\alpha k}(q^4\bar{q})= \mathcal{A}
\left[\psi_3(q_1q_2q_3)\psi_2(q_4\bar{q_5})\right]^{[c]IJ}_{W_cM_IM_J},
\label{phys}
\end{equation}
where ${\cal A}$ is a normalized antisymmetrization operator for
four quarks. $\psi_3$ and $\psi_2$ are wavefunctions of baryon and
meson, respectively. $[~]$ means coupling in terms of the SU$_3^c$,
SU$_2^\tau$, SU$_2^\sigma$ CG coefficients so that it has color
symmetry $[c]W_{c}$, isospin $IM_{I}$, and spin $JM_{J}$.
$\alpha=(YIJ)$, $k$ represents the quantum numbers
$\nu_3,\nu_2,c_3,\cdots,J_2$, which specify the baryon and meson
states. In order to relate the physical bases to the symmetry bases,
the 4-quark cluster basis is introduced
\begin{equation}
\Psi_{\alpha_4 k_4}(q^4) = {\cal A} \left[ \psi_3(q_1q_2q_3)
\psi_1(q_4) \right]^{[c_4 ]I_4J_4}_{W_{c_4}M_{I_4}M_{J_4}},
\label{q4}
\end{equation}
where $\alpha_4=(Y_4I_4J_4)$, $k_4$ represents the quantum numbers
$\nu_3,\nu_1,c_3,\cdots,J_1$. Coupling the anti-quark basis to
4-quark basis Eq.(\ref{q4}) in terms of SU$_3^c$, SU$_2^\tau$,
SU$_2^\sigma$ CG coefficients, gives the 5-quark basis
\[
\left[ \Psi_{\alpha_4 k_4}(q^4))
\psi_{[\bar{c}]\bar{I}\bar{J}}(\bar{q}_5) \right]^{[c]IJ}_{WM_IM_J}
\]
It relates to the physical basis through the SU$_3^c$, SU$_2^\tau$,
SU$_2^\sigma$ Racah coefficients,
\begin{eqnarray}
\Psi_{\alpha k}(q^4\bar{q})&=& U(c_3c_1cc_{\bar{1}};c_4c_2)
U(I_3I_1II_{\bar{1}};I_4I_2)U(J_3J_1JJ_{\bar{1}};J_4J_2)  \nonumber\\
& &\times \left[ \Psi_{\alpha_4 k_4}(q^4))
\psi_{[\bar{c}]\bar{I}\bar{J}}(\bar{q}_5) \right]^{[c]IJ}_{WM_IM_J},
 \label{phys}
\end{eqnarray}
where $U'$s are Racah coefficients, defined as,
\begin{eqnarray}
U(c_3c_1cc_{\bar{1}};c_4c_2)&=&\sum_{\mbox{all }W's}
C^{[c_4]W_{c_4}}_{[c_3]W_{c_3},[c_1]W_{c_1}}
C^{[c]W_c}_{[c_4]W_{c_4},[c_{\bar{1}}]W_{c_{\bar{1}}}}
C^{[c_2]W_{c_2}}_{[c_1]W_{c_1},[c_{\bar{1}}]W_{c_{\bar{1}}}}
C^{[c]W_{c}}_{[c_3]W_{c_3},[c_2]W_{c_2}}. \nonumber
\end{eqnarray}
The Racah coefficients can be found in Ref. \cite{ISFBook}.

The symmetry basis of 4-quark system can be defined as,
\begin{equation}
\Phi_{\alpha_4 K_4}(q^4) = \left| \begin{array}{c} [\nu_4]W_{\nu_4}
\\ \left[ c_4 \right] W_{c_4} [ \mu_4][f_4]Y_4I_4M_{I_4}J_4M_{J_4}
\end{array} \right\rangle,\label{symm_q4}
\end{equation}
which is the basis vector belonging to the irreducible
representations of group chain Eq.(\ref{group-chain}) with $n=4$. In
Eq.(\ref{symm_q4}) $K_4$ stands for $\nu_4\mu_4f_4$. Coupling the
antiquark to Eq.(\ref{symm_q4}), the symmetry bases for 5-quark is
obtained
\begin{equation}
\Phi_{\alpha K}(q^4\bar{q}) = \left[ \Phi_{\alpha_4 K_4}(q^4))
\psi_{[\bar{c}]\bar{I}\bar{J}}(\bar{q}_5) \right]^{[c]IJ}_{WM_IM_J}.
\label{symm_q5}
\end{equation}

To relate the symmetry basis to physical basis, first we express the
four-quark cluster basis Eq.(\ref{q4}) in terms of the symmetry
basis Eq.(\ref{symm_q4}) \cite{Harvey,chen1,qdcsm1},
\begin{eqnarray}
\Psi_{\alpha_4 k_4}(q^4) &=&\sum_{\tilde{\nu_{4}}\mu_{4}f_{4}}
C^{[\tilde{\nu}_4][c_4][\mu_4]}_{[\tilde{\nu}_3][c_3][\mu_3],[\tilde{\nu}_1][c_1][\mu_1]}
C^{[\mu_4][f_4][J_4]}_{[\mu_3][f_3][J_3],[\mu_1][f_1][J_1]}
C^{[f_4]Y_4I_4}_{[f_3]Y_3I_3,[f_1]Y_1I_1}\Phi_{\alpha_4 K_4}(q^4),
\label{trans1}
\end{eqnarray}
$C'$s are the isoscalar factors (ISFs) of
SU$_{mn}\supset$SU$_m\times$SU$_n$, which can be obtained from
Chen's book \cite{ISFBook}. Then, the physical bases and symmetry
bases for 5-quark can be transformed to each other by
\begin{eqnarray}
\Psi_{\alpha k}(q^4\bar{q}) & = & \sum_{K} C_{k K} \Phi_{\alpha
K}(q^4\bar{q}) \nonumber \\
&=&\sum_{\tilde{\nu_{4}}\mu_{4}f_{4}}\{UUU\}_{\mbox{\scriptsize{Racah}}}\{CCC\}_{\mbox{\scriptsize{ISF}}}
\Phi_{\alpha K}(q^4\bar{q}). \label{trans2}
\end{eqnarray}
here, $\{UUU\}_{\mbox{\scriptsize{Racah}}}$ are the Racah
coefficients in Eq. (\ref{phys});
$\{CCC\}_{\mbox{\scriptsize{ISF}}}$ are the isoscalar factors in Eq.
(\ref{trans1}). All the transformation coefficients are tabulated in
the appendix.

A physical 5-quark state with quantum number $\alpha =(YIJ)$ is
expressed as a channel coupling wave function
\begin{equation}
\Psi_{\alpha}(q^4\bar{q}) = \sum_k C_k \Psi_{\alpha k}(q^4\bar{q}).
\end{equation}
The channel coupling coefficients $C_k$ is determined by the
diagonalization of the 5-quark Hamiltonian as usual. As before, the
calculation of Hamiltonian matrix elements in the cluster bases is
replaced by that of matrix elements in the symmetry bases, where the
fractional parentage technique can be used. The calculation of the
matrix elements in symmetry bases is given in Ref.\cite{group1}, we
will not repeat the procedure here.

\section{Quark models and calculation method}

The models used in the calculations are naive quark model, Salamanca
chiral quark model and quark delocalization color screening model.
The Hamiltonian of these models have been given in the first paper
of this series \cite{group1}. The only modification is the
delocalized single particle orbital wavefunction, which is given
below for the baryon-meson basis:
\begin{eqnarray}
\psi_l(\vec{r}) & = & (\phi_L+\epsilon_1\phi_R)/N(\epsilon_1),
 \nonumber \\
\psi_r(\vec{r}) & = & (\phi_R+\epsilon_2\phi_L)/N(\epsilon_2), \\
N(\epsilon_1) & = & \sqrt{1+\epsilon_1^2+2\epsilon_1 \langle
\phi_R|\phi_L \rangle} . \\
 N(\epsilon_2) & = &
\sqrt{1+\epsilon_2^2+2\epsilon_2 \langle \phi_R|\phi_L \rangle} .
\nonumber
\end{eqnarray}
where $\epsilon_{1}$ denotes the quark in baryon orbit delocalized
into meson orbit, while $\epsilon_{2}$ is from meson to baryon. They
are determined by the dynamics of the 5-quark system.

The calculation method is quite the same as in Ref.\cite{group1}.

\section{Results and discussions}

To take into consideration of the effects of various color
structures, the hidden color channels are included in this
calculation, which is different from our dibaryon calculation where
only color singlet channels are included \cite{dibaryon,qdcsm1}.
Like the systematic study of the JW di-quark configuration, all
possible states within the $u$, $d$, $s$ three-flavor world have
been calculated in this baryon-meson configuration. Both
single-channel and channel coupling calculations have been carried
out in three quark models.
To save space, only several general features and comparison between
the JW di-quark and meson-baryon two configurations and among three
models are given below.

(1) Generally there exist effective attractions for almost all the
states both in the extended QDCSM and the chiral quark model, while
there are about ten channels which are pure repulsive in the naive
quark model ($NN$ interaction study already showed that the naive
quark model is not realistic for hadron interaction, it can not
provide the intermediate range $NN$ attraction). In QDCSM, the
attraction between decuplet baryons and vector mesons are usually
large ($> 200$ MeV), the attraction between octet baryons and
pseudo-scalar mesons are small (less than 10 MeV), and the
attraction between octet (decuplet) baryons and vector
(pseudo-scalar) mesons lie between. There are also several
exceptions. This situation is similar to the baryon-baryon
interactions where the attraction between decuplet baryons is large,
between octet baryons is small and between decuplet and octet
baryons is in between. In chiral quark model, the above general
features are kept but with more exceptions. Fig.1 gives several
examples. It is clear that $\Delta \rho, \Delta \bar{K}^*$ and
$\Sigma^* \bar{K}^*$ channels have strong attractions, whereas
$\Sigma\pi, \Sigma K$ and $\Xi\bar{K}$ show pure repulsive or very
small attractions. More examples can be found in Fig.3-6.

\begin{center}
\epsfysize=3.0in \epsfbox{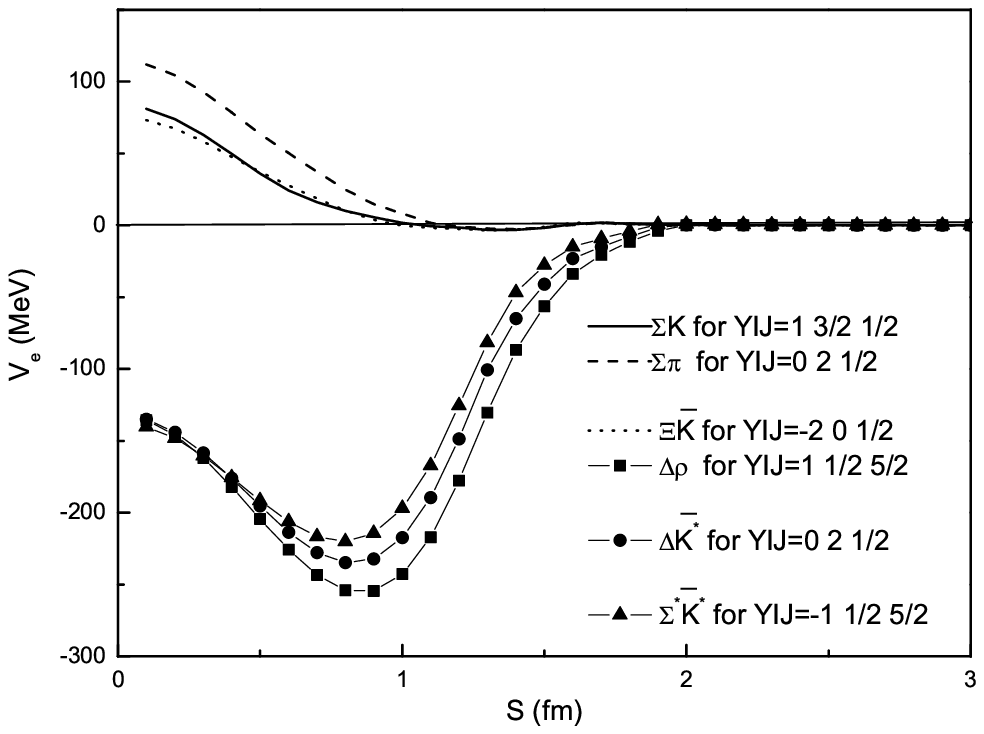}

Fig.1 Effective potentials vs cluster separation in QDCSM.


\end{center}

(2) For chiral quark model and naive quark model, most states have
very similar results (energy difference less than 30 MeV) for two
configurations: $qq$-$qq$-$\bar{q}$ and $qqq$-$q\bar{q}$. For these
states, the separations between quark clusters corresponding to the
minimum-energy are small ($S$ and/or $T < 0.7$ fm). The differences
between two configurations tend to disappear after channel-coupling,
the remaining minor differences come from the different quark
orbital wave functions used. There are also several states, the
energy differences are large (greater than 60 MeV, even 100 MeV) for
two configurations. For these states, the separation between quark
clusters corresponding to minimum-energy in baryon-meson
configuration is a little larger (color-singlet clusters can stay
far away), while the ones in JW di-quark configuration are smaller
(colorful clusters tend to stay close), that makes the energies
different. However, the extended QDCSM gives different results. The
di-quark configuration always gives lower energy than baryon-meson
configuration. The main reason is that the color confinement is
screened between quark pair in different clusters in QDCSM. The more
the number of clusters, the lower the model energy. The screened
color confinement assumed for quark pair in two color singlet
clusters was shown to be reasonable by the previous studies
\cite{dibaryon,qdcsm}, however extends to colorful clusters may be
questionable. One may expect there might be additional energy for
separating color clusters. Further investigation is needed.

(3) For the states in anti-decuplet, 27-plet and 35-plet the mass
hierarchies obtained in meson-baryon configuration are quite similar
to that of JW di-quark configuration. Here only the mass hierarchies
of $8_f + \overline{10}_{f}$ states with $J^{P}=\frac{1}{2}^{+}$ of
two configurations in three quark models are given in Fig.2.
\begin{center}
\epsfxsize=4.0in \epsfbox{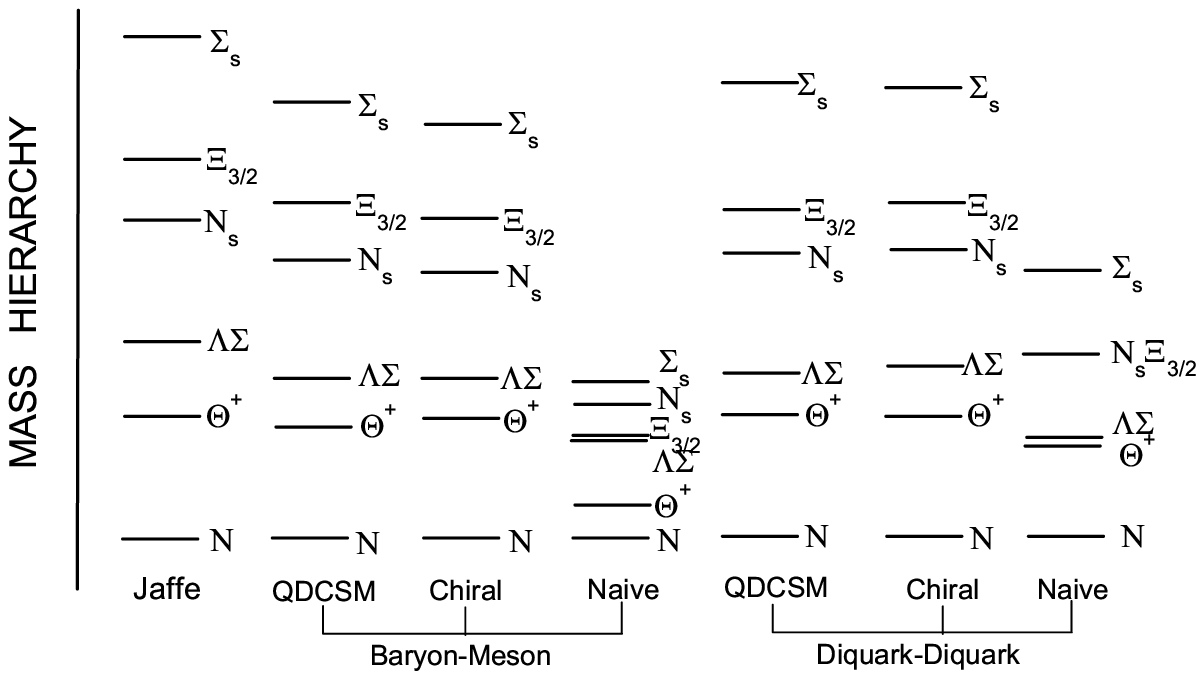}

Fig.2 mass hierarchy in the $8_f + \overline{10}_{f}$ with $J^{P} =
\frac{1}{2}^{+}$ in three quark models, compared with the one in
diquark-diquark configuration and Ref.\cite{JW}.
\end{center}

(4) Figs.3-6 show the effect of channel coupling (sc stands for
single channel, cs for color singlet channel coupling and cc for
full channel coupling). For most channels, the effect of channel
coupling is small, especially for QDCSM. It is worth to note that
the hidden-color states have minor effect on the lowest energy of a
channel. Figs.7-8 show that channel coupling effect for
$(YIJ)=(00\frac{1}{2})$, where the channel coupling contributes a
considerable attraction. However the contribution of hidden-color
channels are still negligible, especially for QDCSM. We like to
emphasize that this result is based on the assumption that the quark
interaction used for hadron spectroscopy and color singlet hadron
channels can be directly extend to hidden color channels and this
assumption is questionable. In fact we don't have any information
about the quark interaction between hidden color channels and
colorless ones up to now.

\begin{center}
\epsfxsize=4.0in \epsfbox{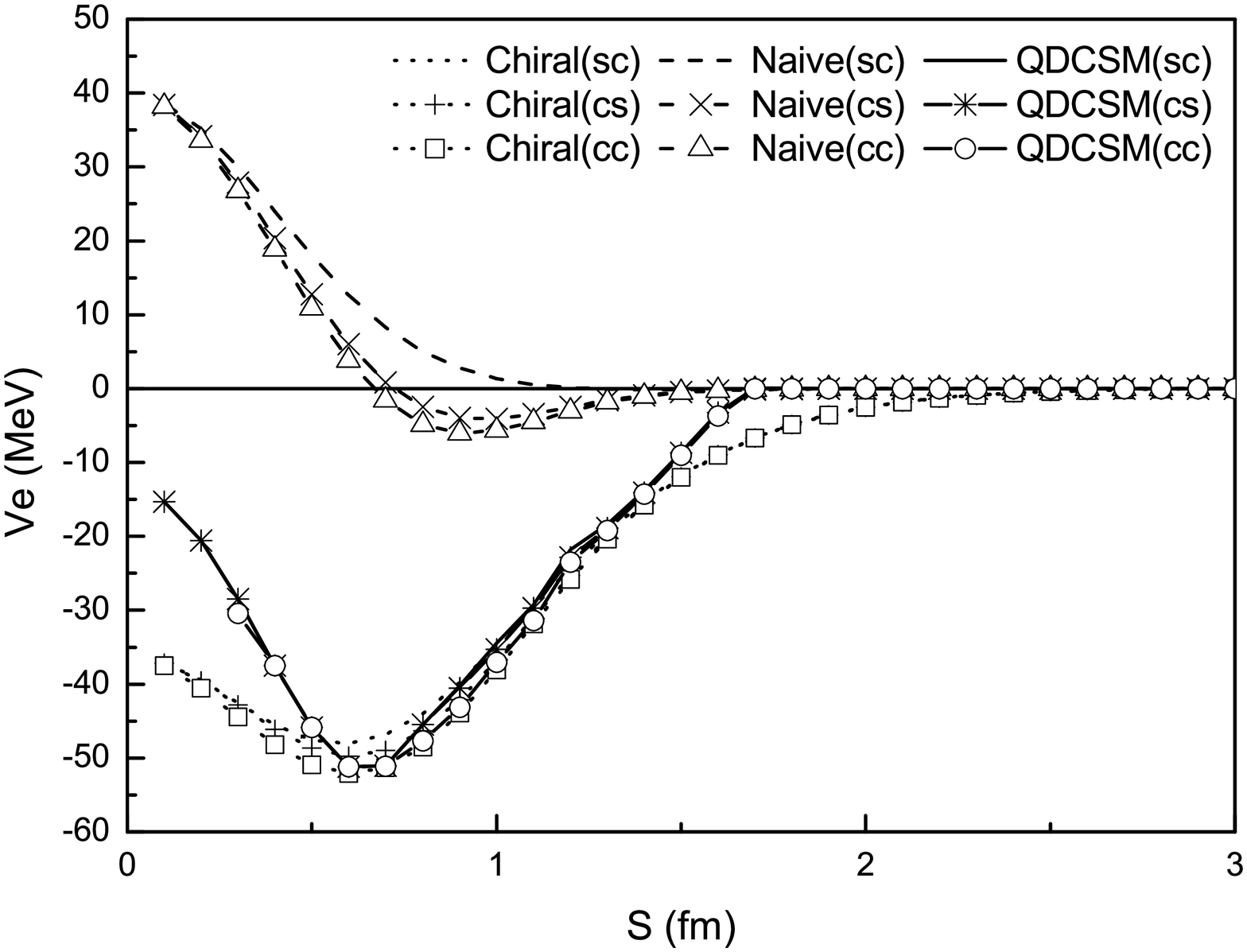}

Fig.3 Effective potentials for $(YIJ)=(20\frac{1}{2})$.

\epsfysize=3.0in \epsfbox{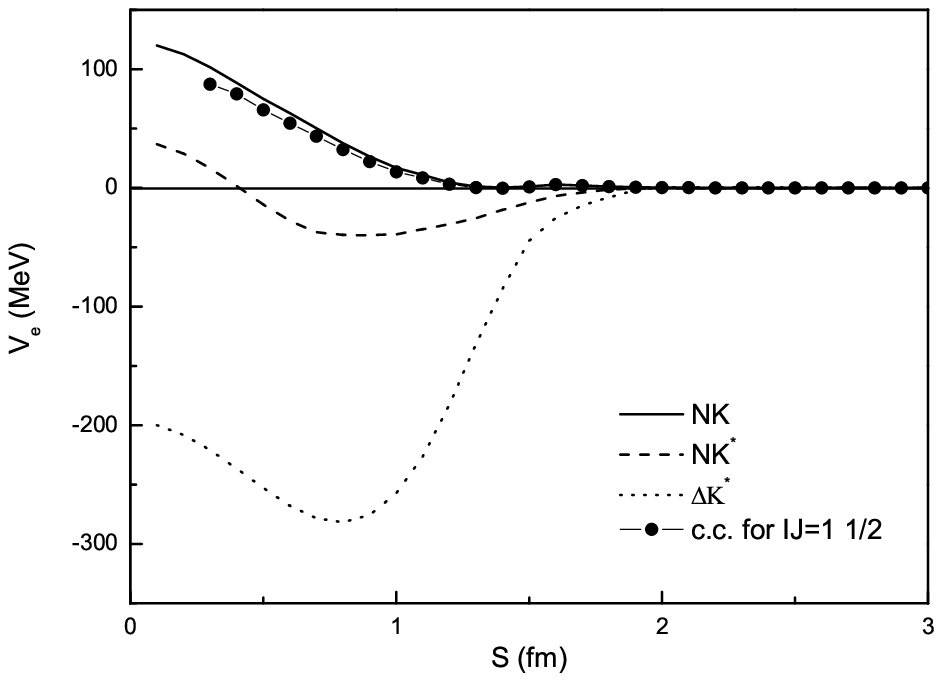}

Fig.4 Effective potentials for $(YIJ)=(21\frac{1}{2})$ in QDCSM.

\epsfysize=3.0in \epsfbox{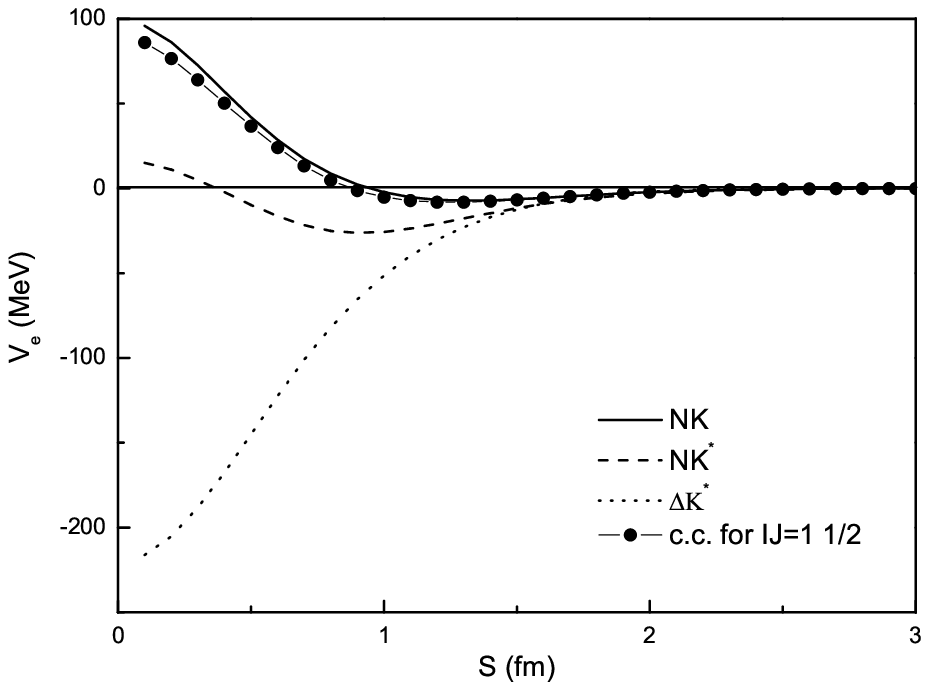}

Fig.5 Same as fig.4 for chiral quark model.

\epsfysize=3.0in \epsfbox{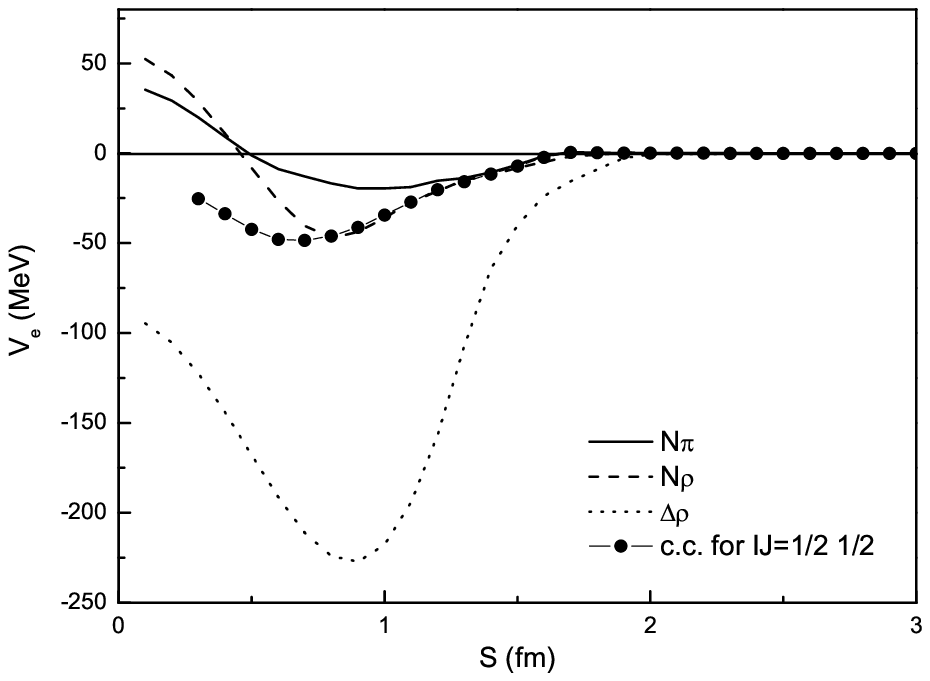}

Fig.6 Effective potentials $(YIJ)=(1\frac{1}{2}\frac{1}{2})$. in
QDCSM.

\epsfysize=3.0in \epsfbox{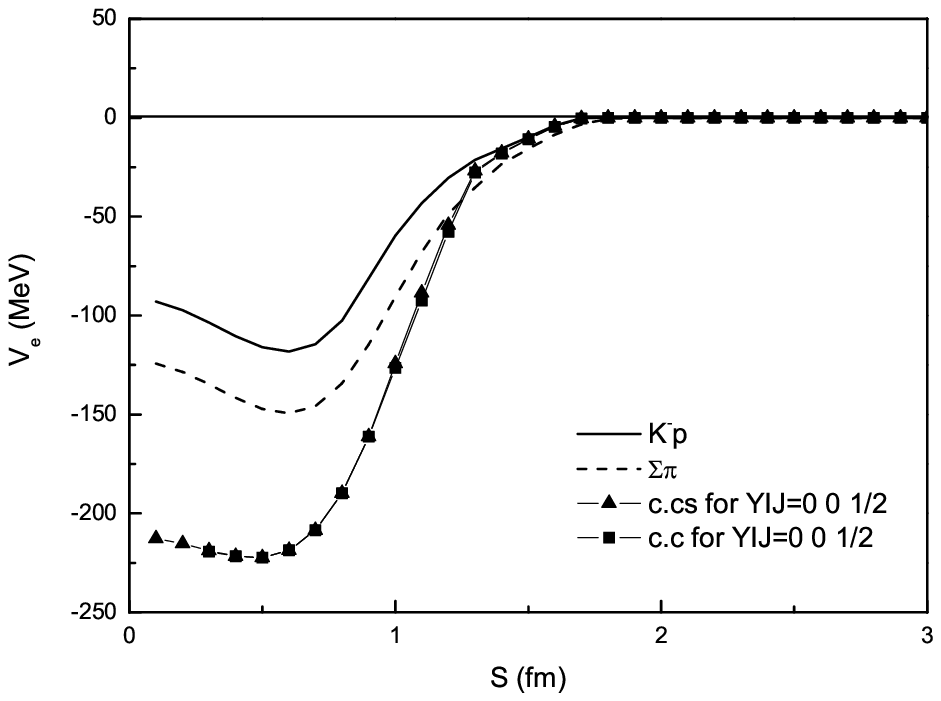}

Fig.7 Effective potentials for $(YIJ)=(00\frac{1}{2})$ in QDCSM.

\epsfysize=3.0in \epsfbox{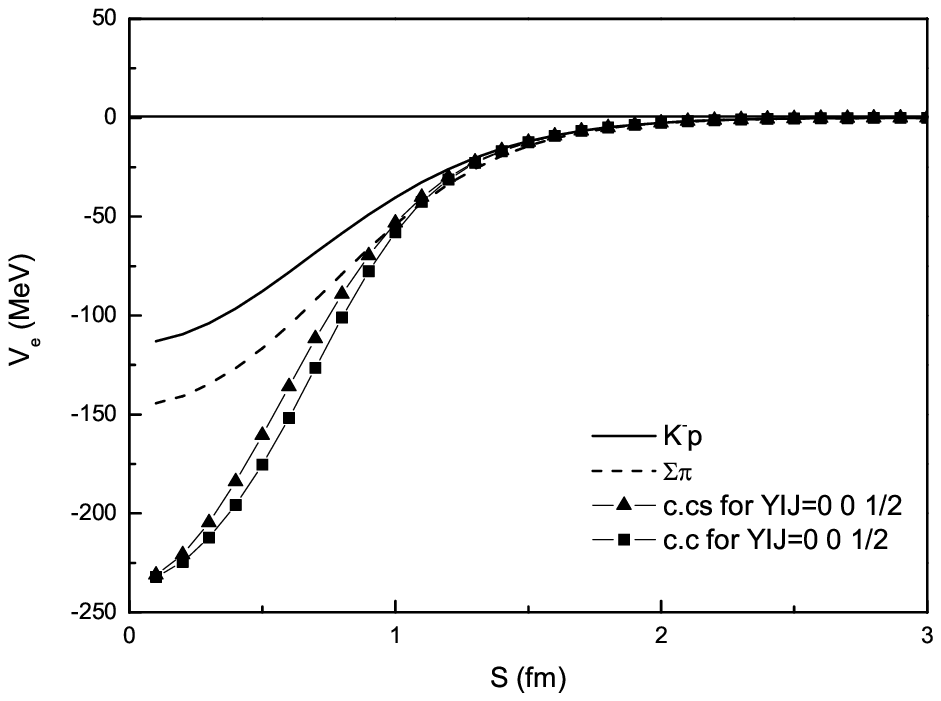}

Fig.8 Effective potentials for $(YIJ)=(00\frac{1}{2})$ in chiral
quark model.
\end{center}



(5) Generally there exist attractions for baryon-meson system,
however in most cases, the attraction is not enough to make the
energy lower than the threshold, i.e., the sum of corresponding
baryon and meson mass. Therefore there will be no narrow resonances
except there is special reason to prevent the decay. Take
$\Theta^{+}$ for example, the lowest energy state here is the $KN$
state. The effective potential for the $N-K$ channels in three quark
models is given in Fig.3. For the $I=0$ S-wave $KN$ state, although
the naive quark model gives repulsive potentials, there is a
moderate attraction $\sim 51$~MeV at a separation of 0.6 fm in the
QDCSM and a little weaker attraction $\sim 48$~MeV at a separation
of 0.6 fm in the chiral quark model. (The results are different from
the previous calculation \cite{Zhang}, where no attraction was
obtained. The reason is the effect of $\sigma$ exchange was reduced
in Ref.\cite{Zhang} by using larger mass and ideal mixing of
$\sigma_0$ and $\sigma_8$). In QDCSM, the mass of $\Theta^{+}$ is
1851 MeV, however the mass of kaon is poorly estimated to be 647~MeV
in QDCSM. If we make a correction of the physical mass of kaon, the
energy of the state can be reduced to 1700~MeV, which is still too
high to match the experimental value 1540 MeV. In the chiral quark
model, the case is quite similar to QDCSM.

(6) Finally, there are several interesting states in QDCSM: (1)
Their masses are lower than the lowest two- and three-body decay
threshold and they might be better pentaquark candidates. However
the results are sensitive to model details. For
$(YIJ)=(1\frac{1}{2}\frac{5}{2})$ channel, the lowest state has mass
around 1822 MeV, which is not only lower than the two-body decay
thresholds ($\Delta\rho$), but also lower than the three-body decay
thresholds ($N\pi\rho$). And for $(YIJ)=(00\frac{5}{2})$ channel,
there is a state with mass around 2017 MeV, which is lower than the
corresponding two-body decay threshold $\Sigma^{*}\rho$ and the
three-body decay thresholds ($\Lambda\pi\rho$). (2) $K^{-}p$ state
has been calculated dynamically in order to see if there is a
resonance state $\Lambda(1405)$. The mass of $K^{-}p$ is 1560 MeV.
If we make a correction of the physical mass of kaon, the energy of
the state can be reduced to 1408~MeV, which corresponds to the
observed mass of $\Lambda(1405)$. (3) In Ref.\cite{Zhang2}, several
quasibound states have been proposed , such as I=1 $\Delta K$,
$I=\frac{1}{2}$ $\Sigma K$, $I=\frac{1}{2},\frac{3}{2}$ $N\phi$, and
so on. In QDCSM, all these states have been calculated and strong
attraction ($> 100$ MeV) were obtained. The dynamical calculation of
these states are in progress.

\section{Summary}
Multi-quark system is complicate, our knowledge about multi-quark
system is limited and there is no well established multi-quark state
experimentally up to now. Nevertheless its study is indispensable if
one wants to understand the properties of the abundant color
structures allowed in QCD. The systematic study could provide the
general features of the multi-quark system. In this case, a powerful
method is needed. This paper (and the previous one\cite{group1})
reports a group theoretic method for five-quark calculation. The
method is applied to a systematic study of all possible five-quark
states within $u,d,s$ three-flavor world in baryon-meson
configuration with three quark models. The powerful feature of group
theory method is shown by the large amount of spectroscopy data
obtained easily in this approach (only very limited results have
been shown in this report).

In the pentaquark studies few years ago various constituent quark
models were proposed. Our transformation between different physical
bases to the same set of symmetry bases shows these models are a
choice of different parts of the model space of the five-quark
system. Our calculation shows that if one uses a large enough model
space the differences of these different configuration choices tend
to disappear.

Besides the five-quark calculation, the baryon-meson scattering is
another field where the technique developed in these two papers can
be used. With the progress on the meson spectroscopy of
non-relativistic quark model \cite{meson}, the unified description
of hadron properties and hadron-hadron interaction becomes possible.
There are a lot of work devoted to baryon-baryon scattering,
baryon-meson scattering is less studied. Employing the group
theoretic method presented here one can study the baryon-meson
scattering within the framework of constituent quark models.

There are experimental indication and theoretical motivation that
the pure $q^3$ baryon and $q\bar{q}$ meson configuration should be
modified to include higher Fock components. The effect of the
five-quark component $q^3q\bar{q}$ of baryon can be studied as well
by the same technique developed here.

This work is supported by the NSFC 10375030, 90503011, 10435080,
10775072.

\section*{Appendix}

Table A1. Transformation coefficients between physical bases and
symmetry bases. The column labels are $[\nu_{4}]$, $[\mu_{4}]$,
$[f_{4}]$, $[\sigma_{J_{4}}]$, $[I_{4}]$. For first four labels, 1
stands for the symmetry $[4]$; 2, $[31]$; 3, $[22]$; 4, $[211]$, and
for the last one, 1 stands for isospin 2; 2, $\frac{3}{2}$; 3, 1; 4,
$\frac{1}{2}$; 5, 0. The row labels are $B$ (baryon index) and $M$
(meson index), where indices $B$ and $M$ are listed in Table A2. The
transformation coefficients should be the square root of the
entries, and a negative sign means to take the negative square root.

\begin{tiny}

\begin{tabular}{c|cccccccccccccccc}  \hline
\multicolumn{ 4}{c}{$Y= 2$~~  $I=2          $~~  $J=\frac{5}{2}$~~
$Y_4=\frac{4}{3} $}  \\  \hline
 B  M & 21111 \\  \hline
 5  7 & $ 1$  \\ \hline
\end{tabular}

\begin{tabular}{c|cccccccccccccccc}  \hline
\multicolumn{ 4}{c}{$Y= 2$~~  $I=2          $~~  $J=\frac{3}{2}$~~
$Y_4=\frac{4}{3} $}  \\  \hline
 B  M & 12121 & 21111 & 22121 \\  \hline
 5  2 & $-\frac{  1}{   8}$ & $ \frac{  5}{   8}$ & $-\frac{  1}{   4}$  \\
 5  7 & $ \frac{  5}{  24}$ & $ \frac{  3}{   8}$ & $ \frac{  5}{  12}$  \\
17 17 & $-\frac{  2}{   3}$ & $ 0$ & $ \frac{  1}{   3}$  \\
\hline
\end{tabular}

\begin{tabular}{c|cccccccccccccccc}  \hline
\multicolumn{ 4}{c}{$Y= 2$~~  $I=2          $~~  $J=\frac{1}{2}$~~
$Y_4=\frac{4}{3} $}  \\  \hline
 B  M & 12121 & 22121 & 23131 \\  \hline
 5  7 & $ \frac{  1}{   3}$ & $ \frac{  2}{   3}$ & $ 0$  \\
17 12 & $-\frac{  1}{   2}$ & $ \frac{  1}{   4}$ & $-\frac{  1}{   4}$  \\
17 17 & $-\frac{  1}{   6}$ & $ \frac{  1}{  12}$ & $ \frac{  3}{
4}$  \\ \hline
\end{tabular}

\begin{tabular}{c|cccccccccccccccc}  \hline
\multicolumn{ 4}{c}{$Y= 2$~~  $I=1          $~~  $J=\frac{5}{2}$~~
$Y_4=\frac{4}{3} $}  \\  \hline
 B  M & 12213 & 22213 \\  \hline
 5  7 & $-\frac{  1}{   3}$ & $-\frac{  2}{   3}$  \\
13 17 & $-\frac{  2}{   3}$ & $ \frac{  1}{   3}$  \\ \hline
\end{tabular}

\begin{tabular}{c|cccccccccccccccc}  \hline
\multicolumn{ 8}{c}{$Y= 2$~~  $I=1          $~~  $J=\frac{3}{2}$~~
$Y_4=\frac{4}{3} $}  \\  \hline
 B  M & 12223 & 21223 & 22223 & 12213 & 22213 & 23223 & 24223 \\  \hline
 1  7 & $ \frac{  1}{   9}$ & $ \frac{  2}{   3}$ & $ \frac{  2}{   9}$ & $ 0$ & $ 0$ & $ 0$ & $ 0$  \\
 5  2 & $-\frac{  1}{  12}$ & $ \frac{  1}{   8}$ & $-\frac{  1}{   6}$ & $-\frac{  5}{  24}$ & $-\frac{  5}{  12}$ & $ 0$ & $ 0$  \\
 5  7 & $ \frac{  5}{  36}$ & $-\frac{  5}{  24}$ & $ \frac{  5}{  18}$ & $-\frac{  1}{   8}$ & $-\frac{  1}{   4}$ & $ 0$ & $ 0$  \\
 9 17 & $-\frac{  4}{   9}$ & $ 0$ & $ \frac{  2}{   9}$ & $ 0$ & $ 0$ & $-\frac{  1}{   3}$ & $ 0$  \\
13 12 & $ \frac{  1}{  24}$ & $ 0$ & $-\frac{  1}{  48}$ & $-\frac{  5}{  12}$ & $ \frac{  5}{  24}$ & $-\frac{  1}{   8}$ & $-\frac{  3}{  16}$  \\
13 17 & $-\frac{  5}{  72}$ & $ 0$ & $ \frac{  5}{ 144}$ & $-\frac{  1}{   4}$ & $ \frac{  1}{   8}$ & $ \frac{  5}{  24}$ & $ \frac{  5}{  16}$  \\
17 17 & $ \frac{  1}{   9}$ & $ 0$ & $-\frac{  1}{  18}$ & $ 0$ &
$ 0$ & $-\frac{  1}{   3}$ & $ \frac{  1}{   2}$  \\ \hline
\end{tabular}

\begin{tabular}{c|cccccccccccccccc}  \hline
\multicolumn{ 9}{c}{$Y= 2$~~  $I=1          $~~  $J=\frac{1}{2}$~~
$Y_4=\frac{4}{3} $}  \\  \hline
 B  M & 12223 & 12233 & 21223 & 22223 & 22233 & 23223 & 24233 & 24223 \\  \hline
 1  2 & $ \frac{  1}{  12}$ & $-\frac{  1}{  12}$ & $ \frac{  1}{   2}$ & $ \frac{  1}{   6}$ & $-\frac{  1}{   6}$ & $ 0$ & $ 0$ & $ 0$  \\
 1  7 & $ \frac{  1}{  36}$ & $ \frac{  1}{   4}$ & $ \frac{  1}{   6}$ & $ \frac{  1}{  18}$ & $ \frac{  1}{   2}$ & $ 0$ & $ 0$ & $ 0$  \\
 5  7 & $ \frac{  2}{   9}$ & $ 0$ & $-\frac{  1}{   3}$ & $ \frac{  4}{   9}$ & $ 0$ & $ 0$ & $ 0$ & $ 0$  \\
 9 12 & $-\frac{  1}{   3}$ & $-\frac{  1}{  12}$ & $ 0$ & $ \frac{  1}{   6}$ & $ \frac{  1}{  24}$ & $-\frac{  1}{   4}$ & $-\frac{  1}{   8}$ & $ 0$  \\
 9 17 & $-\frac{  1}{   9}$ & $ \frac{  1}{   4}$ & $ 0$ & $ \frac{  1}{  18}$ & $-\frac{  1}{   8}$ & $-\frac{  1}{  12}$ & $ \frac{  3}{   8}$ & $ 0$  \\
13 17 & $-\frac{  1}{   9}$ & $ 0$ & $ 0$ & $ \frac{  1}{  18}$ & $ 0$ & $ \frac{  1}{   3}$ & $ 0$ & $ \frac{  1}{   2}$  \\
17 12 & $ \frac{  1}{  12}$ & $-\frac{  1}{  12}$ & $ 0$ & $-\frac{  1}{  24}$ & $ \frac{  1}{  24}$ & $-\frac{  1}{   4}$ & $ \frac{  1}{   8}$ & $ \frac{  3}{   8}$  \\
17 17 & $ \frac{  1}{  36}$ & $ \frac{  1}{   4}$ & $ 0$ &
$-\frac{  1}{  72}$ & $-\frac{  1}{   8}$ & $-\frac{  1}{  12}$ &
$-\frac{  3}{   8}$ & $ \frac{  1}{   8}$  \\ \hline
\end{tabular}

\begin{tabular}{c|cccccccccccccccc}  \hline
\multicolumn{ 4}{c}{$Y= 2$~~  $I=0          $~~  $J=\frac{5}{2}$~~
$Y_4=\frac{4}{3} $}  \\  \hline
 B  M & 23315 \\  \hline
13 17 & $-1$  \\ \hline
\end{tabular}

\begin{tabular}{c|cccccccccccccccc}  \hline
\multicolumn{ 5}{c}{$Y= 2$~~  $I=0          $~~  $J=\frac{3}{2}$~~
$Y_4=\frac{4}{3} $}  \\  \hline
 B  M & 12325 & 22325 & 24325 & 23315 \\  \hline
 1  7 & $-\frac{  1}{   3}$ & $-\frac{  2}{   3}$ & $ 0$ & $ 0$  \\
 9 17 & $-\frac{  1}{   3}$ & $ \frac{  1}{   6}$ & $-\frac{  1}{   2}$ & $ 0$  \\
13 12 & $-\frac{  1}{   8}$ & $ \frac{  1}{  16}$ & $ \frac{  3}{  16}$ & $-\frac{  5}{   8}$  \\
13 17 & $ \frac{  5}{  24}$ & $-\frac{  5}{  48}$ & $-\frac{  5}{
16}$ & $-\frac{  3}{   8}$  \\ \hline
\end{tabular}

\begin{tabular}{c|cccccccccccccccc}  \hline
\multicolumn{ 6}{c}{$Y= 2$~~  $I=0          $~~  $J=\frac{1}{2}$~~
$Y_4=\frac{4}{3} $}  \\  \hline
 B  M & 12325 & 21335 & 22325 & 23335 & 24325 \\  \hline
 1  2 & $-\frac{  1}{   4}$ & $ \frac{  1}{   4}$ & $-\frac{  1}{   2}$ & $ 0$ & $ 0$  \\
 1  7 & $-\frac{  1}{  12}$ & $-\frac{  3}{   4}$ & $-\frac{  1}{   6}$ & $ 0$ & $ 0$  \\
 9 12 & $-\frac{  1}{   4}$ & $ 0$ & $ \frac{  1}{   8}$ & $-\frac{  1}{   4}$ & $-\frac{  3}{   8}$  \\
 9 17 & $-\frac{  1}{  12}$ & $ 0$ & $ \frac{  1}{  24}$ & $ \frac{  3}{   4}$ & $-\frac{  1}{   8}$  \\
13 17 & $ \frac{  1}{   3}$ & $ 0$ & $-\frac{  1}{   6}$ & $ 0$ &
$-\frac{  1}{   2}$  \\ \hline
\end{tabular}

\begin{tabular}{c|cccccccccccccccc}  \hline
\multicolumn{ 4}{c}{$Y= 1$~~  $I=\frac{5}{2}$~~  $J=\frac{5}{2}$~~
$Y_4=\frac{4}{3} $}  \\  \hline
 B  M & 21111 \\  \hline
 5  6 & $-1$  \\ \hline
\end{tabular}

\begin{tabular}{c|cccccccccccccccc}  \hline
\multicolumn{ 4}{c}{$Y= 1$~~  $I=\frac{5}{2}$~~  $J=\frac{3}{2}$~~
$Y_4=\frac{4}{3} $}  \\  \hline
 B  M & 12121 & 21111 & 22121 \\  \hline
 5  1 & $ \frac{  1}{   8}$ & $-\frac{  5}{   8}$ & $ \frac{  1}{   4}$  \\
 5  6 & $-\frac{  5}{  24}$ & $-\frac{  3}{   8}$ & $-\frac{  5}{  12}$  \\
17 16 & $ \frac{  2}{   3}$ & $ 0$ & $-\frac{  1}{   3}$  \\
\hline
\end{tabular}

\begin{tabular}{c|cccccccccccccccc}  \hline
\multicolumn{ 4}{c}{$Y= 1$~~  $I=\frac{5}{2}$~~  $J=\frac{1}{2}$~~
$Y_4=\frac{4}{3} $}  \\  \hline
 B  M & 12121 & 22121 & 23131 \\  \hline
 5  6 & $-\frac{  1}{   3}$ & $-\frac{  2}{   3}$ & $ 0$  \\
17 11 & $ \frac{  1}{   2}$ & $-\frac{  1}{   4}$ & $ \frac{  1}{   4}$  \\
17 16 & $ \frac{  1}{   6}$ & $-\frac{  1}{  12}$ & $-\frac{  3}{
4}$  \\ \hline
\end{tabular}

\begin{tabular}{c|cccccccccccccccc}  \hline
\multicolumn{ 4}{c}{$Y= 1$~~  $I=\frac{3}{2}$~~  $J=\frac{5}{2}$~~
$Y_4=\frac{4}{3} $}  \\  \hline
 B  M & 12213 & 21111 & 22213 \\  \hline
 5  6 & $ \frac{  5}{  24}$ & $ \frac{  3}{   8}$ & $ \frac{  5}{  12}$  \\
 5  9 & $-\frac{  1}{   8}$ & $ \frac{  5}{   8}$ & $-\frac{  1}{   4}$  \\
13 16 & $ \frac{  2}{   3}$ & $ 0$ & $-\frac{  1}{   3}$  \\
\hline
\end{tabular}

\begin{tabular}{c|cccccccccccccccc}  \hline
\multicolumn{ 4}{c}{$Y= 1$~~  $I=\frac{3}{2}$~~  $J=\frac{5}{2}$~~
$Y_4=\frac{1}{3} $}  \\  \hline
 B  M & 12212 & 21112 & 22212 \\  \hline
 5 10 & $ \frac{  1}{   4}$ & $-\frac{  1}{   4}$ & $ \frac{  1}{   2}$  \\
 6  7 & $-\frac{  1}{  12}$ & $-\frac{  3}{   4}$ & $-\frac{  1}{   6}$  \\
15 17 & $ \frac{  2}{   3}$ & $ 0$ & $-\frac{  1}{   3}$  \\
\hline
\end{tabular}

\begin{tabular}{c|cccccccccccccccc}  \hline
\multicolumn{11}{c}{$Y= 1$~~  $I=\frac{3}{2}$~~  $J=\frac{3}{2}$~~
$Y_4=\frac{4}{3} $}  \\  \hline
 B  M & 12223 & 21223 & 22223 & 12121 & 12213 & 21111 & 22121 & 22213 & 23223 & 24223 \\  \hline
 1  6 & $-\frac{  1}{   9}$ & $-\frac{  2}{   3}$ & $-\frac{  2}{   9}$ & $ 0$ & $ 0$ & $ 0$ & $ 0$ & $ 0$ & $ 0$ & $ 0$  \\
 5  1 & $ \frac{  5}{  96}$ & $-\frac{  5}{  64}$ & $ \frac{  5}{  48}$ & $-\frac{  3}{  64}$ & $ \frac{ 25}{ 192}$ & $ \frac{ 15}{  64}$ & $-\frac{  3}{  32}$ & $ \frac{ 25}{  96}$ & $ 0$ & $ 0$  \\
 5  5 & $-\frac{  1}{  32}$ & $ \frac{  3}{  64}$ & $-\frac{  1}{  16}$ & $-\frac{  5}{  64}$ & $-\frac{  5}{  64}$ & $ \frac{ 25}{  64}$ & $-\frac{  5}{  32}$ & $-\frac{  5}{  32}$ & $ 0$ & $ 0$  \\
 5  6 & $-\frac{ 25}{ 288}$ & $ \frac{ 25}{ 192}$ & $-\frac{ 25}{ 144}$ & $ \frac{  5}{  64}$ & $ \frac{  5}{  64}$ & $ \frac{  9}{  64}$ & $ \frac{  5}{  32}$ & $ \frac{  5}{  32}$ & $ 0$ & $ 0$  \\
 5  9 & $ \frac{  5}{  96}$ & $-\frac{  5}{  64}$ & $ \frac{  5}{  48}$ & $ \frac{ 25}{ 192}$ & $-\frac{  3}{  64}$ & $ \frac{ 15}{  64}$ & $ \frac{ 25}{  96}$ & $-\frac{  3}{  32}$ & $ 0$ & $ 0$  \\
 9 16 & $ \frac{  4}{   9}$ & $ 0$ & $-\frac{  2}{   9}$ & $ 0$ & $ 0$ & $ 0$ & $ 0$ & $ 0$ & $ \frac{  1}{   3}$ & $ 0$  \\
13 11 & $-\frac{  1}{  24}$ & $ 0$ & $ \frac{  1}{  48}$ & $ 0$ & $ \frac{  5}{  12}$ & $ 0$ & $ 0$ & $-\frac{  5}{  24}$ & $ \frac{  1}{   8}$ & $ \frac{  3}{  16}$  \\
13 16 & $ \frac{  5}{  72}$ & $ 0$ & $-\frac{  5}{ 144}$ & $ 0$ & $ \frac{  1}{   4}$ & $ 0$ & $ 0$ & $-\frac{  1}{   8}$ & $-\frac{  5}{  24}$ & $-\frac{  5}{  16}$  \\
17 16 & $-\frac{  5}{  72}$ & $ 0$ & $ \frac{  5}{ 144}$ & $-\frac{  1}{   4}$ & $ 0$ & $ 0$ & $ \frac{  1}{   8}$ & $ 0$ & $ \frac{  5}{  24}$ & $-\frac{  5}{  16}$  \\
17 19 & $ \frac{  1}{  24}$ & $ 0$ & $-\frac{  1}{  48}$ &
$-\frac{  5}{  12}$ & $ 0$ & $ 0$ & $ \frac{  5}{  24}$ & $ 0$ &
$-\frac{  1}{   8}$ & $ \frac{  3}{  16}$  \\ \hline
\end{tabular}

\begin{tabular}{c|cccccccccccccccc}  \hline
\multicolumn{11}{c}{$Y= 1$~~  $I=\frac{3}{2}$~~  $J=\frac{3}{2}$~~
$Y_4=\frac{1}{3} $}  \\  \hline
 B  M & 12222 & 21222 & 22222 & 12122 & 12212 & 21112 & 22122 & 22212 & 23222 & 24222 \\  \hline
 3  7 & $-\frac{  1}{   9}$ & $-\frac{  2}{   3}$ & $-\frac{  2}{   9}$ & $ 0$ & $ 0$ & $ 0$ & $ 0$ & $ 0$ & $ 0$ & $ 0$  \\
 5  4 & $ \frac{  1}{  16}$ & $-\frac{  3}{  32}$ & $ \frac{  1}{   8}$ & $ \frac{  1}{  32}$ & $ \frac{  5}{  32}$ & $-\frac{  5}{  32}$ & $ \frac{  1}{  16}$ & $ \frac{  5}{  16}$ & $ 0$ & $ 0$  \\
 5 10 & $-\frac{  5}{  48}$ & $ \frac{  5}{  32}$ & $-\frac{  5}{  24}$ & $-\frac{  5}{  96}$ & $ \frac{  3}{  32}$ & $-\frac{  3}{  32}$ & $-\frac{  5}{  48}$ & $ \frac{  3}{  16}$ & $ 0$ & $ 0$  \\
 6  2 & $-\frac{  1}{  48}$ & $ \frac{  1}{  32}$ & $-\frac{  1}{  24}$ & $ \frac{  3}{  32}$ & $-\frac{  5}{  96}$ & $-\frac{ 15}{  32}$ & $ \frac{  3}{  16}$ & $-\frac{  5}{  48}$ & $ 0$ & $ 0$  \\
 6  7 & $ \frac{  5}{ 144}$ & $-\frac{  5}{  96}$ & $ \frac{  5}{  72}$ & $-\frac{  5}{  32}$ & $-\frac{  1}{  32}$ & $-\frac{  9}{  32}$ & $-\frac{  5}{  16}$ & $-\frac{  1}{  16}$ & $ 0$ & $ 0$  \\
11 17 & $ \frac{  4}{   9}$ & $ 0$ & $-\frac{  2}{   9}$ & $ 0$ & $ 0$ & $ 0$ & $ 0$ & $ 0$ & $ \frac{  1}{   3}$ & $ 0$  \\
15 12 & $-\frac{  1}{  24}$ & $ 0$ & $ \frac{  1}{  48}$ & $ 0$ & $ \frac{  5}{  12}$ & $ 0$ & $ 0$ & $-\frac{  5}{  24}$ & $ \frac{  1}{   8}$ & $ \frac{  3}{  16}$  \\
15 17 & $ \frac{  5}{  72}$ & $ 0$ & $-\frac{  5}{ 144}$ & $ 0$ & $ \frac{  1}{   4}$ & $ 0$ & $ 0$ & $-\frac{  1}{   8}$ & $-\frac{  5}{  24}$ & $-\frac{  5}{  16}$  \\
17 20 & $-\frac{  1}{  12}$ & $ 0$ & $ \frac{  1}{  24}$ & $ \frac{  1}{   6}$ & $ 0$ & $ 0$ & $-\frac{  1}{  12}$ & $ 0$ & $ \frac{  1}{   4}$ & $-\frac{  3}{   8}$  \\
18 17 & $ \frac{  1}{  36}$ & $ 0$ & $-\frac{  1}{  72}$ & $
\frac{  1}{   2}$ & $ 0$ & $ 0$ & $-\frac{  1}{   4}$ & $ 0$ &
$-\frac{  1}{  12}$ & $ \frac{  1}{   8}$  \\ \hline
\end{tabular}

\begin{tabular}{c|cccccccccccccccc}  \hline
\multicolumn{12}{c}{$Y= 1$~~  $I=\frac{3}{2}$~~  $J=\frac{1}{2}$~~
$Y_4=\frac{4}{3} $}  \\  \hline
 B  M & 12223 & 12233 & 21223 & 22223 & 22233 & 12121 & 22121 & 23223 & 24233 & 24223 & 23131 \\  \hline
 1  1 & $-\frac{  1}{  12}$ & $ \frac{  1}{  12}$ & $-\frac{  1}{   2}$ & $-\frac{  1}{   6}$ & $ \frac{  1}{   6}$ & $ 0$ & $ 0$ & $ 0$ & $ 0$ & $ 0$ & $ 0$  \\
 1  6 & $-\frac{  1}{  36}$ & $-\frac{  1}{   4}$ & $-\frac{  1}{   6}$ & $-\frac{  1}{  18}$ & $-\frac{  1}{   2}$ & $ 0$ & $ 0$ & $ 0$ & $ 0$ & $ 0$ & $ 0$  \\
 5  6 & $-\frac{  5}{  36}$ & $ 0$ & $ \frac{  5}{  24}$ & $-\frac{  5}{  18}$ & $ 0$ & $ \frac{  1}{   8}$ & $ \frac{  1}{   4}$ & $ 0$ & $ 0$ & $ 0$ & $ 0$  \\
 5  9 & $ \frac{  1}{  12}$ & $ 0$ & $-\frac{  1}{   8}$ & $ \frac{  1}{   6}$ & $ 0$ & $ \frac{  5}{  24}$ & $ \frac{  5}{  12}$ & $ 0$ & $ 0$ & $ 0$ & $ 0$  \\
 9 11 & $ \frac{  1}{   3}$ & $ \frac{  1}{  12}$ & $ 0$ & $-\frac{  1}{   6}$ & $-\frac{  1}{  24}$ & $ 0$ & $ 0$ & $ \frac{  1}{   4}$ & $ \frac{  1}{   8}$ & $ 0$ & $ 0$  \\
 9 16 & $ \frac{  1}{   9}$ & $-\frac{  1}{   4}$ & $ 0$ & $-\frac{  1}{  18}$ & $ \frac{  1}{   8}$ & $ 0$ & $ 0$ & $ \frac{  1}{  12}$ & $-\frac{  3}{   8}$ & $ 0$ & $ 0$  \\
\hline
\end{tabular}

\begin{tabular}{c|cccccccccccccccc}  \hline
\multicolumn{12}{c}{$Y= 1$~~  $I=\frac{3}{2}$~~  $J=\frac{1}{2}$~~
$Y_4=\frac{1}{3} $}  \\  \hline
 B  M & 12222 & 12232 & 21222 & 22222 & 22232 & 12122 & 22122 & 23222 & 24232 & 24222 & 23132 \\  \hline
 3  2 & $-\frac{  1}{  12}$ & $ \frac{  1}{  12}$ & $-\frac{  1}{   2}$ & $-\frac{  1}{   6}$ & $ \frac{  1}{   6}$ & $ 0$ & $ 0$ & $ 0$ & $ 0$ & $ 0$ & $ 0$  \\
 3  7 & $-\frac{  1}{  36}$ & $-\frac{  1}{   4}$ & $-\frac{  1}{   6}$ & $-\frac{  1}{  18}$ & $-\frac{  1}{   2}$ & $ 0$ & $ 0$ & $ 0$ & $ 0$ & $ 0$ & $ 0$  \\
 5 10 & $-\frac{  1}{   6}$ & $ 0$ & $ \frac{  1}{   4}$ & $-\frac{  1}{   3}$ & $ 0$ & $-\frac{  1}{  12}$ & $-\frac{  1}{   6}$ & $ 0$ & $ 0$ & $ 0$ & $ 0$  \\
 6  7 & $ \frac{  1}{  18}$ & $ 0$ & $-\frac{  1}{  12}$ & $ \frac{  1}{   9}$ & $ 0$ & $-\frac{  1}{   4}$ & $-\frac{  1}{   2}$ & $ 0$ & $ 0$ & $ 0$ & $ 0$  \\
11 12 & $ \frac{  1}{   3}$ & $ \frac{  1}{  12}$ & $ 0$ & $-\frac{  1}{   6}$ & $-\frac{  1}{  24}$ & $ 0$ & $ 0$ & $ \frac{  1}{   4}$ & $ \frac{  1}{   8}$ & $ 0$ & $ 0$  \\
11 17 & $ \frac{  1}{   9}$ & $-\frac{  1}{   4}$ & $ 0$ & $-\frac{  1}{  18}$ & $ \frac{  1}{   8}$ & $ 0$ & $ 0$ & $ \frac{  1}{  12}$ & $-\frac{  3}{   8}$ & $ 0$ & $ 0$  \\
\hline
\end{tabular}

\begin{tabular}{c|cccccccccccccccc}  \hline
\multicolumn{ 4}{c}{$Y= 1$~~  $I=\frac{1}{2}$~~  $J=\frac{5}{2}$~~
$Y_4=\frac{4}{3} $}  \\  \hline
 B  M & 12213 & 22213 & 23315 \\  \hline
 5  6 & $-\frac{  1}{   3}$ & $-\frac{  2}{   3}$ & $ 0$  \\
13 16 & $-\frac{  1}{   6}$ & $ \frac{  1}{  12}$ & $ \frac{  3}{   4}$  \\
13 19 & $-\frac{  1}{   2}$ & $ \frac{  1}{   4}$ & $-\frac{  1}{
4}$  \\ \hline
\end{tabular}

\begin{tabular}{c|cccccccccccccccc}  \hline
\multicolumn{ 5}{c}{$Y= 1$~~  $I=\frac{1}{2}$~~  $J=\frac{5}{2}$~~
$Y_4=\frac{1}{3} $}  \\  \hline
 B  M & 12214 & 22214 & 23314 & 24414 \\  \hline
 6  7 & $ \frac{  1}{   3}$ & $ \frac{  2}{   3}$ & $ 0$ & $ 0$  \\
13 20 & $ \frac{  1}{   4}$ & $-\frac{  1}{   8}$ & $ \frac{  1}{   4}$ & $-\frac{  3}{   8}$  \\
14 17 & $ \frac{  3}{   8}$ & $-\frac{  3}{  16}$ & $-\frac{  3}{   8}$ & $ \frac{  1}{  16}$  \\
15 17 & $ \frac{  1}{  24}$ & $-\frac{  1}{  48}$ & $ \frac{  3}{
8}$ & $ \frac{  9}{  16}$  \\ \hline
\end{tabular}

\begin{tabular}{c|cccccccccccccccc}  \hline
\multicolumn{12}{c}{$Y= 1$~~  $I=\frac{1}{2}$~~  $J=\frac{3}{2}$~~
$Y_4=\frac{4}{3} $}  \\  \hline
 B  M & 12223 & 12325 & 21223 & 22223 & 22325 & 12213 & 22213 & 23223 & 24325 & 23315 & 24223 \\  \hline
 1  6 & $ \frac{  1}{  36}$ & $ \frac{  1}{   4}$ & $ \frac{  1}{   6}$ & $ \frac{  1}{  18}$ & $ \frac{  1}{   2}$ & $ 0$ & $ 0$ & $ 0$ & $ 0$ & $ 0$ & $ 0$  \\
 1  9 & $ \frac{  1}{  12}$ & $-\frac{  1}{  12}$ & $ \frac{  1}{   2}$ & $ \frac{  1}{   6}$ & $-\frac{  1}{   6}$ & $ 0$ & $ 0$ & $ 0$ & $ 0$ & $ 0$ & $ 0$  \\
 5  1 & $-\frac{  1}{  12}$ & $ 0$ & $ \frac{  1}{   8}$ & $-\frac{  1}{   6}$ & $ 0$ & $-\frac{  5}{  24}$ & $-\frac{  5}{  12}$ & $ 0$ & $ 0$ & $ 0$ & $ 0$  \\
 5  6 & $ \frac{  5}{  36}$ & $ 0$ & $-\frac{  5}{  24}$ & $ \frac{  5}{  18}$ & $ 0$ & $-\frac{  1}{   8}$ & $-\frac{  1}{   4}$ & $ 0$ & $ 0$ & $ 0$ & $ 0$  \\
 9 16 & $-\frac{  1}{   9}$ & $ \frac{  1}{   4}$ & $ 0$ & $ \frac{  1}{  18}$ & $-\frac{  1}{   8}$ & $ 0$ & $ 0$ & $-\frac{  1}{  12}$ & $ \frac{  3}{   8}$ & $ 0$ & $ 0$  \\
 9 19 & $-\frac{  1}{   3}$ & $-\frac{  1}{  12}$ & $ 0$ & $ \frac{  1}{   6}$ & $ \frac{  1}{  24}$ & $ 0$ & $ 0$ & $-\frac{  1}{   4}$ & $-\frac{  1}{   8}$ & $ 0$ & $ 0$  \\
\hline
\end{tabular}

\begin{tabular}{c|cccccccccccccccc}  \hline
\multicolumn{17}{c}{$Y= 1$~~  $I=\frac{1}{2}$~~  $J=\frac{3}{2}$~~
$Y_4=\frac{1}{3} $}  \\  \hline
 B  M & 12224 & 12324 & 12424 & 21224 & 22224 & 22324 & 22424 & 12214 & 22214 & 23224 & 23424 & 24324 & 24424 & 23314 & 24224 & 24414 \\  \hline
 1 10 & $-\frac{  1}{  24}$ & $ \frac{  1}{  12}$ & $ \frac{  1}{   8}$ & $-\frac{  1}{   4}$ & $-\frac{  1}{  12}$ & $ \frac{  1}{   6}$ & $ \frac{  1}{   4}$ & $ 0$ & $ 0$ & $ 0$ & $ 0$ & $ 0$ & $ 0$ & $ 0$ & $ 0$ & $ 0$  \\
 2  7 & $-\frac{  1}{  16}$ & $-\frac{  1}{   8}$ & $-\frac{  1}{  48}$ & $-\frac{  3}{   8}$ & $-\frac{  1}{   8}$ & $-\frac{  1}{   4}$ & $-\frac{  1}{  24}$ & $ 0$ & $ 0$ & $ 0$ & $ 0$ & $ 0$ & $ 0$ & $ 0$ & $ 0$ & $ 0$  \\
 3  7 & $-\frac{  1}{ 144}$ & $ \frac{  1}{   8}$ & $-\frac{  3}{  16}$ & $-\frac{  1}{  24}$ & $-\frac{  1}{  72}$ & $ \frac{  1}{   4}$ & $-\frac{  3}{   8}$ & $ 0$ & $ 0$ & $ 0$ & $ 0$ & $ 0$ & $ 0$ & $ 0$ & $ 0$ & $ 0$  \\
 6  2 & $ \frac{  1}{  12}$ & $ 0$ & $ 0$ & $-\frac{  1}{   8}$ & $ \frac{  1}{   6}$ & $ 0$ & $ 0$ & $ \frac{  5}{  24}$ & $ \frac{  5}{  12}$ & $ 0$ & $ 0$ & $ 0$ & $ 0$ & $ 0$ & $ 0$ & $ 0$  \\
 6  7 & $-\frac{  5}{  36}$ & $ 0$ & $ 0$ & $ \frac{  5}{  24}$ & $-\frac{  5}{  18}$ & $ 0$ & $ 0$ & $ \frac{  1}{   8}$ & $ \frac{  1}{   4}$ & $ 0$ & $ 0$ & $ 0$ & $ 0$ & $ 0$ & $ 0$ & $ 0$  \\
 9 20 & $ \frac{  1}{   6}$ & $ \frac{  1}{  12}$ & $ 0$ & $ 0$ & $-\frac{  1}{  12}$ & $-\frac{  1}{  24}$ & $ 0$ & $ 0$ & $ 0$ & $ \frac{  1}{   8}$ & $-\frac{  1}{   8}$ & $ \frac{  1}{   8}$ & $-\frac{  1}{   4}$ & $ 0$ & $ 0$ & $ 0$  \\
\hline
\end{tabular}

\begin{tabular}{c|cccccccccccccccc}  \hline
\multicolumn{14}{c}{$Y= 1$~~  $I=\frac{1}{2}$~~  $J=\frac{1}{2}$~~
$Y_4=\frac{4}{3} $}  \\  \hline
 B  M & 12223 & 12233 & 12325 & 21223 & 21335 & 22223 & 22233 & 22325 & 23223 & 23335 & 24233 & 24325 & 24223 \\  \hline
 1  1 & $ \frac{  1}{  48}$ & $-\frac{  1}{  48}$ & $ \frac{  3}{  16}$ & $ \frac{  1}{   8}$ & $-\frac{  3}{  16}$ & $ \frac{  1}{  24}$ & $-\frac{  1}{  24}$ & $ \frac{  3}{   8}$ & $ 0$ & $ 0$ & $ 0$ & $ 0$ & $ 0$  \\
 1  5 & $ \frac{  1}{  16}$ & $-\frac{  1}{  16}$ & $-\frac{  1}{  16}$ & $ \frac{  3}{   8}$ & $ \frac{  1}{  16}$ & $ \frac{  1}{   8}$ & $-\frac{  1}{   8}$ & $-\frac{  1}{   8}$ & $ 0$ & $ 0$ & $ 0$ & $ 0$ & $ 0$  \\
 1  6 & $ \frac{  1}{ 144}$ & $ \frac{  1}{  16}$ & $ \frac{  1}{  16}$ & $ \frac{  1}{  24}$ & $ \frac{  9}{  16}$ & $ \frac{  1}{  72}$ & $ \frac{  1}{   8}$ & $ \frac{  1}{   8}$ & $ 0$ & $ 0$ & $ 0$ & $ 0$ & $ 0$  \\
 1  9 & $ \frac{  1}{  48}$ & $ \frac{  3}{  16}$ & $-\frac{  1}{  48}$ & $ \frac{  1}{   8}$ & $-\frac{  3}{  16}$ & $ \frac{  1}{  24}$ & $ \frac{  3}{   8}$ & $-\frac{  1}{  24}$ & $ 0$ & $ 0$ & $ 0$ & $ 0$ & $ 0$  \\
 5  6 & $ \frac{  2}{   9}$ & $ 0$ & $ 0$ & $-\frac{  1}{   3}$ & $ 0$ & $ \frac{  4}{   9}$ & $ 0$ & $ 0$ & $ 0$ & $ 0$ & $ 0$ & $ 0$ & $ 0$  \\
 9 11 & $-\frac{  1}{  12}$ & $-\frac{  1}{  48}$ & $ \frac{  3}{  16}$ & $ 0$ & $ 0$ & $ \frac{  1}{  24}$ & $ \frac{  1}{  96}$ & $-\frac{  3}{  32}$ & $-\frac{  1}{  16}$ & $ \frac{  3}{  16}$ & $-\frac{  1}{  32}$ & $ \frac{  9}{  32}$ & $ 0$  \\
\hline
\end{tabular}

\begin{tabular}{c|cccccccccccccccc}  \hline
\multicolumn{17}{c}{$Y= 1$~~  $I=\frac{1}{2}$~~  $J=\frac{1}{2}$~~
$Y_4=\frac{1}{3} $}  \\  \hline
 B  M & 12224 & 12234 & 12324 & 12424 & 12434 & 21224 & 21334 & 22224 & 22234 & 22324 & 22424 & 22434 & 23224 & 23334 & 23424 & 24234 \\  \hline
 1  4 & $-\frac{  1}{  32}$ & $ \frac{  1}{  32}$ & $ \frac{  1}{  16}$ & $ \frac{  3}{  32}$ & $ \frac{  1}{  32}$ & $-\frac{  3}{  16}$ & $-\frac{  1}{  16}$ & $-\frac{  1}{  16}$ & $ \frac{  1}{  16}$ & $ \frac{  1}{   8}$ & $ \frac{  3}{  16}$ & $ \frac{  1}{  16}$ & $ 0$ & $ 0$ & $ 0$ & $ 0$  \\
 1 10 & $-\frac{  1}{  96}$ & $-\frac{  3}{  32}$ & $ \frac{  1}{  48}$ & $ \frac{  1}{  32}$ & $-\frac{  3}{  32}$ & $-\frac{  1}{  16}$ & $ \frac{  3}{  16}$ & $-\frac{  1}{  48}$ & $-\frac{  3}{  16}$ & $ \frac{  1}{  24}$ & $ \frac{  1}{  16}$ & $-\frac{  3}{  16}$ & $ 0$ & $ 0$ & $ 0$ & $ 0$  \\
 2  2 & $-\frac{  3}{  64}$ & $ \frac{  3}{  64}$ & $-\frac{  3}{  32}$ & $-\frac{  1}{  64}$ & $-\frac{  1}{ 192}$ & $-\frac{  9}{  32}$ & $ \frac{  3}{  32}$ & $-\frac{  3}{  32}$ & $ \frac{  3}{  32}$ & $-\frac{  3}{  16}$ & $-\frac{  1}{  32}$ & $-\frac{  1}{  96}$ & $ 0$ & $ 0$ & $ 0$ & $ 0$  \\
 2  7 & $-\frac{  1}{  64}$ & $-\frac{  9}{  64}$ & $-\frac{  1}{  32}$ & $-\frac{  1}{ 192}$ & $ \frac{  1}{  64}$ & $-\frac{  3}{  32}$ & $-\frac{  9}{  32}$ & $-\frac{  1}{  32}$ & $-\frac{  9}{  32}$ & $-\frac{  1}{  16}$ & $-\frac{  1}{  96}$ & $ \frac{  1}{  32}$ & $ 0$ & $ 0$ & $ 0$ & $ 0$  \\
 3  2 & $-\frac{  1}{ 192}$ & $ \frac{  1}{ 192}$ & $ \frac{  3}{  32}$ & $-\frac{  9}{  64}$ & $-\frac{  3}{  64}$ & $-\frac{  1}{  32}$ & $-\frac{  3}{  32}$ & $-\frac{  1}{  96}$ & $ \frac{  1}{  96}$ & $ \frac{  3}{  16}$ & $-\frac{  9}{  32}$ & $-\frac{  3}{  32}$ & $ 0$ & $ 0$ & $ 0$ & $ 0$  \\
\hline
\end{tabular}

\begin{tabular}{c|cccccccccccccccc}  \hline
 B  M & 24324 & 24424 & 24434 & 24224 \\  \hline
 1  4 & $ 0$ & $ 0$ & $ 0$ & $ 0$  \\
 1 10 & $ 0$ & $ 0$ & $ 0$ & $ 0$  \\
 2  2 & $ 0$ & $ 0$ & $ 0$ & $ 0$  \\
 2  7 & $ 0$ & $ 0$ & $ 0$ & $ 0$  \\
 3  2 & $ 0$ & $ 0$ & $ 0$ & $ 0$  \\
 3  7 & $ 0$ & $ 0$ & $ 0$ & $ 0$  \\
 6  7 & $ 0$ & $ 0$ & $ 0$ & $ 0$  \\
 9 14 & $ \frac{  3}{  32}$ & $-\frac{  3}{  16}$ & $-\frac{  3}{  64}$ & $ 0$  \\
 9 20 & $ \frac{  1}{  32}$ & $-\frac{  1}{  16}$ & $ \frac{  9}{  64}$ & $ 0$  \\
10 12 & $-\frac{  9}{  64}$ & $ \frac{  1}{  32}$ & $ \frac{  1}{ 128}$ & $ 0$  \\
10 17 & $-\frac{  3}{  64}$ & $ \frac{  1}{  96}$ & $-\frac{  3}{ 128}$ & $ 0$  \\
11 12 & $ \frac{  9}{  64}$ & $ \frac{  9}{  32}$ & $ \frac{  9}{ 128}$ & $ 0$  \\
11 17 & $ \frac{  3}{  64}$ & $ \frac{  3}{  32}$ & $-\frac{ 27}{ 128}$ & $ 0$  \\
13 20 & $ \frac{  1}{   8}$ & $ \frac{  1}{  16}$ & $ 0$ & $-\frac{  3}{  16}$  \\
14 17 & $-\frac{  3}{  16}$ & $-\frac{  1}{  96}$ & $ 0$ & $-\frac{  9}{  32}$  \\
15 17 & $ \frac{  3}{  16}$ & $-\frac{  3}{  32}$ & $ 0$ & $-\frac{  1}{  32}$  \\
18 12 & $ 0$ & $ 0$ & $ 0$ & $-\frac{  3}{   8}$  \\
18 17 & $ 0$ & $ 0$ & $ 0$ & $-\frac{  1}{   8}$  \\
21 12 & $ 0$ & $-\frac{  1}{   8}$ & $ \frac{  1}{   8}$ & $ 0$  \\
21 17 & $ 0$ & $-\frac{  1}{  24}$ & $-\frac{  3}{   8}$ & $ 0$
\\ \hline
\end{tabular}

\begin{tabular}{c|cccccccccccccccc}  \hline
\multicolumn{ 4}{c}{$Y= 0$~~  $I=2          $~~  $J=\frac{5}{2}$~~
$Y_4=\frac{1}{3} $}  \\  \hline
 B  M & 12212 & 21112 & 22212 \\  \hline
 5  8 & $-\frac{  1}{   4}$ & $ \frac{  1}{   4}$ & $-\frac{  1}{   2}$  \\
 6  6 & $ \frac{  1}{  12}$ & $ \frac{  3}{   4}$ & $ \frac{  1}{   6}$  \\
15 16 & $-\frac{  2}{   3}$ & $ 0$ & $ \frac{  1}{   3}$  \\
\hline
\end{tabular}

\begin{tabular}{c|cccccccccccccccc}  \hline
\multicolumn{11}{c}{$Y= 0$~~  $I=2          $~~  $J=\frac{3}{2}$~~
$Y_4=\frac{1}{3} $}  \\  \hline
 B  M & 12222 & 21222 & 22222 & 12122 & 12212 & 21112 & 22122 & 22212 & 23222 & 24222 \\  \hline
 3  6 & $ \frac{  1}{   9}$ & $ \frac{  2}{   3}$ & $ \frac{  2}{   9}$ & $ 0$ & $ 0$ & $ 0$ & $ 0$ & $ 0$ & $ 0$ & $ 0$  \\
 5  3 & $-\frac{  1}{  16}$ & $ \frac{  3}{  32}$ & $-\frac{  1}{   8}$ & $-\frac{  1}{  32}$ & $-\frac{  5}{  32}$ & $ \frac{  5}{  32}$ & $-\frac{  1}{  16}$ & $-\frac{  5}{  16}$ & $ 0$ & $ 0$  \\
 5  8 & $ \frac{  5}{  48}$ & $-\frac{  5}{  32}$ & $ \frac{  5}{  24}$ & $ \frac{  5}{  96}$ & $-\frac{  3}{  32}$ & $ \frac{  3}{  32}$ & $ \frac{  5}{  48}$ & $-\frac{  3}{  16}$ & $ 0$ & $ 0$  \\
 6  1 & $ \frac{  1}{  48}$ & $-\frac{  1}{  32}$ & $ \frac{  1}{  24}$ & $-\frac{  3}{  32}$ & $ \frac{  5}{  96}$ & $ \frac{ 15}{  32}$ & $-\frac{  3}{  16}$ & $ \frac{  5}{  48}$ & $ 0$ & $ 0$  \\
 6  6 & $-\frac{  5}{ 144}$ & $ \frac{  5}{  96}$ & $-\frac{  5}{  72}$ & $ \frac{  5}{  32}$ & $ \frac{  1}{  32}$ & $ \frac{  9}{  32}$ & $ \frac{  5}{  16}$ & $ \frac{  1}{  16}$ & $ 0$ & $ 0$  \\
11 16 & $-\frac{  4}{   9}$ & $ 0$ & $ \frac{  2}{   9}$ & $ 0$ & $ 0$ & $ 0$ & $ 0$ & $ 0$ & $-\frac{  1}{   3}$ & $ 0$  \\
15 11 & $ \frac{  1}{  24}$ & $ 0$ & $-\frac{  1}{  48}$ & $ 0$ & $-\frac{  5}{  12}$ & $ 0$ & $ 0$ & $ \frac{  5}{  24}$ & $-\frac{  1}{   8}$ & $-\frac{  3}{  16}$  \\
15 16 & $-\frac{  5}{  72}$ & $ 0$ & $ \frac{  5}{ 144}$ & $ 0$ & $-\frac{  1}{   4}$ & $ 0$ & $ 0$ & $ \frac{  1}{   8}$ & $ \frac{  5}{  24}$ & $ \frac{  5}{  16}$  \\
17 18 & $ \frac{  1}{  12}$ & $ 0$ & $-\frac{  1}{  24}$ & $-\frac{  1}{   6}$ & $ 0$ & $ 0$ & $ \frac{  1}{  12}$ & $ 0$ & $-\frac{  1}{   4}$ & $ \frac{  3}{   8}$  \\
18 16 & $-\frac{  1}{  36}$ & $ 0$ & $ \frac{  1}{  72}$ &
$-\frac{  1}{   2}$ & $ 0$ & $ 0$ & $ \frac{  1}{   4}$ & $ 0$ & $
\frac{  1}{  12}$ & $-\frac{  1}{   8}$  \\ \hline
\end{tabular}

\begin{tabular}{c|cccccccccccccccc}  \hline
\multicolumn{12}{c}{$Y= 0$~~  $I=2          $~~  $J=\frac{1}{2}$~~
$Y_4=\frac{1}{3} $}  \\  \hline
 B  M & 12222 & 12232 & 21222 & 22222 & 22232 & 12122 & 22122 & 23222 & 24232 & 24222 & 23132 \\  \hline
 3  1 & $ \frac{  1}{  12}$ & $-\frac{  1}{  12}$ & $ \frac{  1}{   2}$ & $ \frac{  1}{   6}$ & $-\frac{  1}{   6}$ & $ 0$ & $ 0$ & $ 0$ & $ 0$ & $ 0$ & $ 0$  \\
 3  6 & $ \frac{  1}{  36}$ & $ \frac{  1}{   4}$ & $ \frac{  1}{   6}$ & $ \frac{  1}{  18}$ & $ \frac{  1}{   2}$ & $ 0$ & $ 0$ & $ 0$ & $ 0$ & $ 0$ & $ 0$  \\
 5  8 & $ \frac{  1}{   6}$ & $ 0$ & $-\frac{  1}{   4}$ & $ \frac{  1}{   3}$ & $ 0$ & $ \frac{  1}{  12}$ & $ \frac{  1}{   6}$ & $ 0$ & $ 0$ & $ 0$ & $ 0$  \\
 6  6 & $-\frac{  1}{  18}$ & $ 0$ & $ \frac{  1}{  12}$ & $-\frac{  1}{   9}$ & $ 0$ & $ \frac{  1}{   4}$ & $ \frac{  1}{   2}$ & $ 0$ & $ 0$ & $ 0$ & $ 0$  \\
11 11 & $-\frac{  1}{   3}$ & $-\frac{  1}{  12}$ & $ 0$ & $ \frac{  1}{   6}$ & $ \frac{  1}{  24}$ & $ 0$ & $ 0$ & $-\frac{  1}{   4}$ & $-\frac{  1}{   8}$ & $ 0$ & $ 0$  \\
11 16 & $-\frac{  1}{   9}$ & $ \frac{  1}{   4}$ & $ 0$ & $ \frac{  1}{  18}$ & $-\frac{  1}{   8}$ & $ 0$ & $ 0$ & $-\frac{  1}{  12}$ & $ \frac{  3}{   8}$ & $ 0$ & $ 0$  \\
\hline
\end{tabular}

\begin{tabular}{c|cccccccccccccccc}  \hline
\multicolumn{ 8}{c}{$Y= 0$~~  $I=1          $~~  $J=\frac{5}{2}$~~
$Y_4=\frac{1}{3} $}  \\  \hline
 B  M & 12212 & 21112 & 22212 & 12214 & 22214 & 23314 & 24414 \\  \hline
 5  8 & $ \frac{  1}{   4}$ & $-\frac{  1}{   4}$ & $ \frac{  1}{   2}$ & $ 0$ & $ 0$ & $ 0$ & $ 0$  \\
 6  6 & $-\frac{  1}{  36}$ & $-\frac{  1}{   4}$ & $-\frac{  1}{  18}$ & $-\frac{  2}{   9}$ & $-\frac{  4}{   9}$ & $ 0$ & $ 0$  \\
 6  9 & $-\frac{  1}{  18}$ & $-\frac{  1}{   2}$ & $-\frac{  1}{   9}$ & $ \frac{  1}{   9}$ & $ \frac{  2}{   9}$ & $ 0$ & $ 0$  \\
13 18 & $ 0$ & $ 0$ & $ 0$ & $-\frac{  1}{   4}$ & $ \frac{  1}{   8}$ & $-\frac{  1}{   4}$ & $ \frac{  3}{   8}$  \\
14 16 & $ 0$ & $ 0$ & $ 0$ & $-\frac{  3}{   8}$ & $ \frac{  3}{  16}$ & $ \frac{  3}{   8}$ & $-\frac{  1}{  16}$  \\
15 16 & $ \frac{  2}{   9}$ & $ 0$ & $-\frac{  1}{   9}$ & $-\frac{  1}{  36}$ & $ \frac{  1}{  72}$ & $-\frac{  1}{   4}$ & $-\frac{  3}{   8}$  \\
15 19 & $ \frac{  4}{   9}$ & $ 0$ & $-\frac{  2}{   9}$ & $
\frac{  1}{  72}$ & $-\frac{  1}{ 144}$ & $ \frac{  1}{   8}$ & $
\frac{  3}{  16}$  \\ \hline
\end{tabular}

\begin{tabular}{c|cccccccccccccccc}  \hline
\multicolumn{ 5}{c}{$Y= 0$~~  $I=1          $~~  $J=\frac{5}{2}$~~
$Y_4=-\frac{2}{3}$}  \\  \hline
 B  M & 12213 & 21113 & 22213 & 23313 \\  \hline
 6 10 & $-\frac{  1}{   6}$ & $ \frac{  1}{   2}$ & $-\frac{  1}{   3}$ & $ 0$  \\
 7  7 & $ \frac{  1}{   6}$ & $ \frac{  1}{   2}$ & $ \frac{  1}{   3}$ & $ 0$  \\
15 20 & $-\frac{  1}{   3}$ & $ 0$ & $ \frac{  1}{   6}$ & $ \frac{  1}{   2}$  \\
16 17 & $-\frac{  1}{   3}$ & $ 0$ & $ \frac{  1}{   6}$ &
$-\frac{  1}{   2}$  \\ \hline
\end{tabular}

\begin{tabular}{c|cccccccccccccccc}  \hline
\multicolumn{17}{c}{$Y= 0$~~  $I=1          $~~  $J=\frac{3}{2}$~~
$Y_4=\frac{1}{3} $}  \\  \hline
 B  M & 12224 & 12324 & 12424 & 21224 & 22224 & 22324 & 22424 & 12222 & 21222 & 22222 & 12122 & 12212 & 21112 & 22122 & 22212 & 12214 \\  \hline
 1  8 & $ \frac{  1}{  24}$ & $-\frac{  1}{  12}$ & $-\frac{  1}{   8}$ & $ \frac{  1}{   4}$ & $ \frac{  1}{  12}$ & $-\frac{  1}{   6}$ & $-\frac{  1}{   4}$ & $ 0$ & $ 0$ & $ 0$ & $ 0$ & $ 0$ & $ 0$ & $ 0$ & $ 0$ & $ 0$  \\
 2  6 & $ \frac{  1}{  16}$ & $ \frac{  1}{   8}$ & $ \frac{  1}{  48}$ & $ \frac{  3}{   8}$ & $ \frac{  1}{   8}$ & $ \frac{  1}{   4}$ & $ \frac{  1}{  24}$ & $ 0$ & $ 0$ & $ 0$ & $ 0$ & $ 0$ & $ 0$ & $ 0$ & $ 0$ & $ 0$  \\
 3  6 & $ \frac{  1}{ 216}$ & $-\frac{  1}{  12}$ & $ \frac{  1}{   8}$ & $ \frac{  1}{  36}$ & $ \frac{  1}{ 108}$ & $-\frac{  1}{   6}$ & $ \frac{  1}{   4}$ & $-\frac{  1}{  27}$ & $-\frac{  2}{   9}$ & $-\frac{  2}{  27}$ & $ 0$ & $ 0$ & $ 0$ & $ 0$ & $ 0$ & $ 0$  \\
 3  9 & $-\frac{  1}{ 432}$ & $ \frac{  1}{  24}$ & $-\frac{  1}{  16}$ & $-\frac{  1}{  72}$ & $-\frac{  1}{ 216}$ & $ \frac{  1}{  12}$ & $-\frac{  1}{   8}$ & $-\frac{  2}{  27}$ & $-\frac{  4}{   9}$ & $-\frac{  4}{  27}$ & $ 0$ & $ 0$ & $ 0$ & $ 0$ & $ 0$ & $ 0$  \\
 5  3 & $ 0$ & $ 0$ & $ 0$ & $ 0$ & $ 0$ & $ 0$ & $ 0$ & $ \frac{  1}{  16}$ & $-\frac{  3}{  32}$ & $ \frac{  1}{   8}$ & $ \frac{  1}{  32}$ & $ \frac{  5}{  32}$ & $-\frac{  5}{  32}$ & $ \frac{  1}{  16}$ & $ \frac{  5}{  16}$ & $ 0$  \\
 5  8 & $ 0$ & $ 0$ & $ 0$ & $ 0$ & $ 0$ & $ 0$ & $ 0$ & $-\frac{  5}{  48}$ & $ \frac{  5}{  32}$ & $-\frac{  5}{  24}$ & $-\frac{  5}{  96}$ & $ \frac{  3}{  32}$ & $-\frac{  3}{  32}$ & $-\frac{  5}{  48}$ & $ \frac{  3}{  16}$ & $ 0$  \\
\hline
\end{tabular}

\begin{tabular}{c|cccccccccccccccc}  \hline
 B  M & 22214 & 23224 & 23424 & 24324 & 24424 & 23222 & 23314 & 24224 & 24414 & 24222 \\  \hline
 1  8 & $ 0$ & $ 0$ & $ 0$ & $ 0$ & $ 0$ & $ 0$ & $ 0$ & $ 0$ & $ 0$ & $ 0$  \\
 2  6 & $ 0$ & $ 0$ & $ 0$ & $ 0$ & $ 0$ & $ 0$ & $ 0$ & $ 0$ & $ 0$ & $ 0$  \\
 3  6 & $ 0$ & $ 0$ & $ 0$ & $ 0$ & $ 0$ & $ 0$ & $ 0$ & $ 0$ & $ 0$ & $ 0$  \\
 3  9 & $ 0$ & $ 0$ & $ 0$ & $ 0$ & $ 0$ & $ 0$ & $ 0$ & $ 0$ & $ 0$ & $ 0$  \\
 5  3 & $ 0$ & $ 0$ & $ 0$ & $ 0$ & $ 0$ & $ 0$ & $ 0$ & $ 0$ & $ 0$ & $ 0$  \\
 5  8 & $ 0$ & $ 0$ & $ 0$ & $ 0$ & $ 0$ & $ 0$ & $ 0$ & $ 0$ & $ 0$ & $ 0$  \\
 6  1 & $-\frac{  5}{  18}$ & $ 0$ & $ 0$ & $ 0$ & $ 0$ & $ 0$ & $ 0$ & $ 0$ & $ 0$ & $ 0$  \\
\hline
\end{tabular}

\begin{tabular}{c|cccccccccccccccc}  \hline
\multicolumn{15}{c}{$Y= 0$~~  $I=1          $~~  $J=\frac{3}{2}$~~
$Y_4=-\frac{2}{3}$}  \\  \hline
 B  M & 12223 & 12323 & 21223 & 22223 & 22323 & 12123 & 12213 & 21113 & 22123 & 22213 & 23223 & 24323 & 23313 & 24223 \\  \hline
 3 10 & $ \frac{  1}{  18}$ & $ \frac{  1}{   6}$ & $ \frac{  1}{   3}$ & $ \frac{  1}{   9}$ & $ \frac{  1}{   3}$ & $ 0$ & $ 0$ & $ 0$ & $ 0$ & $ 0$ & $ 0$ & $ 0$ & $ 0$ & $ 0$  \\
 4  7 & $ \frac{  1}{  18}$ & $-\frac{  1}{   6}$ & $ \frac{  1}{   3}$ & $ \frac{  1}{   9}$ & $-\frac{  1}{   3}$ & $ 0$ & $ 0$ & $ 0$ & $ 0$ & $ 0$ & $ 0$ & $ 0$ & $ 0$ & $ 0$  \\
 6  4 & $-\frac{  1}{  24}$ & $ 0$ & $ \frac{  1}{  16}$ & $-\frac{  1}{  12}$ & $ 0$ & $-\frac{  1}{  16}$ & $-\frac{  5}{  48}$ & $ \frac{  5}{  16}$ & $-\frac{  1}{   8}$ & $-\frac{  5}{  24}$ & $ 0$ & $ 0$ & $ 0$ & $ 0$  \\
 6 10 & $ \frac{  5}{  72}$ & $ 0$ & $-\frac{  5}{  48}$ & $ \frac{  5}{  36}$ & $ 0$ & $ \frac{  5}{  48}$ & $-\frac{  1}{  16}$ & $ \frac{  3}{  16}$ & $ \frac{  5}{  24}$ & $-\frac{  1}{   8}$ & $ 0$ & $ 0$ & $ 0$ & $ 0$  \\
 7  2 & $ \frac{  1}{  24}$ & $ 0$ & $-\frac{  1}{  16}$ & $ \frac{  1}{  12}$ & $ 0$ & $-\frac{  1}{  16}$ & $ \frac{  5}{  48}$ & $ \frac{  5}{  16}$ & $-\frac{  1}{   8}$ & $ \frac{  5}{  24}$ & $ 0$ & $ 0$ & $ 0$ & $ 0$  \\
 7  7 & $-\frac{  5}{  72}$ & $ 0$ & $ \frac{  5}{  48}$ & $-\frac{  5}{  36}$ & $ 0$ & $ \frac{  5}{  48}$ & $ \frac{  1}{  16}$ & $ \frac{  3}{  16}$ & $ \frac{  5}{  24}$ & $ \frac{  1}{   8}$ & $ 0$ & $ 0$ & $ 0$ & $ 0$  \\
\hline
\end{tabular}

\begin{tabular}{c|cccccccccccccccc}  \hline
\multicolumn{17}{c}{$Y= 0$~~  $I=1          $~~  $J=\frac{1}{2}$~~
$Y_4=\frac{1}{3} $}  \\  \hline
 B  M & 12224 & 12234 & 12324 & 12424 & 12434 & 21224 & 21334 & 22224 & 22234 & 22324 & 22424 & 22434 & 12222 & 12232 & 21222 & 22222 \\  \hline
 1  3 & $ \frac{  1}{  32}$ & $-\frac{  1}{  32}$ & $-\frac{  1}{  16}$ & $-\frac{  3}{  32}$ & $-\frac{  1}{  32}$ & $ \frac{  3}{  16}$ & $ \frac{  1}{  16}$ & $ \frac{  1}{  16}$ & $-\frac{  1}{  16}$ & $-\frac{  1}{   8}$ & $-\frac{  3}{  16}$ & $-\frac{  1}{  16}$ & $ 0$ & $ 0$ & $ 0$ & $ 0$  \\
 1  8 & $ \frac{  1}{  96}$ & $ \frac{  3}{  32}$ & $-\frac{  1}{  48}$ & $-\frac{  1}{  32}$ & $ \frac{  3}{  32}$ & $ \frac{  1}{  16}$ & $-\frac{  3}{  16}$ & $ \frac{  1}{  48}$ & $ \frac{  3}{  16}$ & $-\frac{  1}{  24}$ & $-\frac{  1}{  16}$ & $ \frac{  3}{  16}$ & $ 0$ & $ 0$ & $ 0$ & $ 0$  \\
 2  1 & $ \frac{  3}{  64}$ & $-\frac{  3}{  64}$ & $ \frac{  3}{  32}$ & $ \frac{  1}{  64}$ & $ \frac{  1}{ 192}$ & $ \frac{  9}{  32}$ & $-\frac{  3}{  32}$ & $ \frac{  3}{  32}$ & $-\frac{  3}{  32}$ & $ \frac{  3}{  16}$ & $ \frac{  1}{  32}$ & $ \frac{  1}{  96}$ & $ 0$ & $ 0$ & $ 0$ & $ 0$  \\
 2  6 & $ \frac{  1}{  64}$ & $ \frac{  9}{  64}$ & $ \frac{  1}{  32}$ & $ \frac{  1}{ 192}$ & $-\frac{  1}{  64}$ & $ \frac{  3}{  32}$ & $ \frac{  9}{  32}$ & $ \frac{  1}{  32}$ & $ \frac{  9}{  32}$ & $ \frac{  1}{  16}$ & $ \frac{  1}{  96}$ & $-\frac{  1}{  32}$ & $ 0$ & $ 0$ & $ 0$ & $ 0$  \\
 3  1 & $ \frac{  1}{ 288}$ & $-\frac{  1}{ 288}$ & $-\frac{  1}{  16}$ & $ \frac{  3}{  32}$ & $ \frac{  1}{  32}$ & $ \frac{  1}{  48}$ & $ \frac{  1}{  16}$ & $ \frac{  1}{ 144}$ & $-\frac{  1}{ 144}$ & $-\frac{  1}{   8}$ & $ \frac{  3}{  16}$ & $ \frac{  1}{  16}$ & $-\frac{  1}{  36}$ & $ \frac{  1}{  36}$ & $-\frac{  1}{   6}$ & $-\frac{  1}{  18}$  \\
\hline
\end{tabular}

\begin{tabular}{c|cccccccccccccccc}  \hline
 B  M & 22232 & 12122 & 22122 & 23224 & 23334 & 23424 & 24234 & 24324 & 24424 & 24434 & 23222 & 24232 & 24224 & 24222 & 23132 \\  \hline
 1  3 & $ 0$ & $ 0$ & $ 0$ & $ 0$ & $ 0$ & $ 0$ & $ 0$ & $ 0$ & $ 0$ & $ 0$ & $ 0$ & $ 0$ & $ 0$ & $ 0$ & $ 0$  \\
 1  8 & $ 0$ & $ 0$ & $ 0$ & $ 0$ & $ 0$ & $ 0$ & $ 0$ & $ 0$ & $ 0$ & $ 0$ & $ 0$ & $ 0$ & $ 0$ & $ 0$ & $ 0$  \\
 2  1 & $ 0$ & $ 0$ & $ 0$ & $ 0$ & $ 0$ & $ 0$ & $ 0$ & $ 0$ & $ 0$ & $ 0$ & $ 0$ & $ 0$ & $ 0$ & $ 0$ & $ 0$  \\
 2  6 & $ 0$ & $ 0$ & $ 0$ & $ 0$ & $ 0$ & $ 0$ & $ 0$ & $ 0$ & $ 0$ & $ 0$ & $ 0$ & $ 0$ & $ 0$ & $ 0$ & $ 0$  \\
 3  1 & $ \frac{  1}{  18}$ & $ 0$ & $ 0$ & $ 0$ & $ 0$ & $ 0$ & $ 0$ & $ 0$ & $ 0$ & $ 0$ & $ 0$ & $ 0$ & $ 0$ & $ 0$ & $ 0$  \\
 3  5 & $ \frac{  1}{   9}$ & $ 0$ & $ 0$ & $ 0$ & $ 0$ & $ 0$ & $ 0$ & $ 0$ & $ 0$ & $ 0$ & $ 0$ & $ 0$ & $ 0$ & $ 0$ & $ 0$  \\
 3  6 & $-\frac{  1}{   6}$ & $ 0$ & $ 0$ & $ 0$ & $ 0$ & $ 0$ & $ 0$ & $ 0$ & $ 0$ & $ 0$ & $ 0$ & $ 0$ & $ 0$ & $ 0$ & $ 0$  \\
\hline
\end{tabular}

\begin{tabular}{c|cccccccccccccccc}  \hline
\multicolumn{17}{c}{$Y= 0$~~  $I=1          $~~  $J=\frac{1}{2}$~~
$Y_4=-\frac{2}{3}$}  \\  \hline
 B  M & 12223 & 12233 & 12323 & 21223 & 21333 & 22223 & 22233 & 22323 & 12123 & 22123 & 23223 & 23333 & 24233 & 24323 & 24223 & 23133 \\  \hline
 3  5 & $ \frac{  1}{  24}$ & $-\frac{  1}{  24}$ & $ \frac{  1}{   8}$ & $ \frac{  1}{   4}$ & $-\frac{  1}{   8}$ & $ \frac{  1}{  12}$ & $-\frac{  1}{  12}$ & $ \frac{  1}{   4}$ & $ 0$ & $ 0$ & $ 0$ & $ 0$ & $ 0$ & $ 0$ & $ 0$ & $ 0$  \\
 3 10 & $ \frac{  1}{  72}$ & $ \frac{  1}{   8}$ & $ \frac{  1}{  24}$ & $ \frac{  1}{  12}$ & $ \frac{  3}{   8}$ & $ \frac{  1}{  36}$ & $ \frac{  1}{   4}$ & $ \frac{  1}{  12}$ & $ 0$ & $ 0$ & $ 0$ & $ 0$ & $ 0$ & $ 0$ & $ 0$ & $ 0$  \\
 4  2 & $ \frac{  1}{  24}$ & $-\frac{  1}{  24}$ & $-\frac{  1}{   8}$ & $ \frac{  1}{   4}$ & $ \frac{  1}{   8}$ & $ \frac{  1}{  12}$ & $-\frac{  1}{  12}$ & $-\frac{  1}{   4}$ & $ 0$ & $ 0$ & $ 0$ & $ 0$ & $ 0$ & $ 0$ & $ 0$ & $ 0$  \\
 4  7 & $ \frac{  1}{  72}$ & $ \frac{  1}{   8}$ & $-\frac{  1}{  24}$ & $ \frac{  1}{  12}$ & $-\frac{  3}{   8}$ & $ \frac{  1}{  36}$ & $ \frac{  1}{   4}$ & $-\frac{  1}{  12}$ & $ 0$ & $ 0$ & $ 0$ & $ 0$ & $ 0$ & $ 0$ & $ 0$ & $ 0$  \\
 6 10 & $ \frac{  1}{   9}$ & $ 0$ & $ 0$ & $-\frac{  1}{   6}$ & $ 0$ & $ \frac{  2}{   9}$ & $ 0$ & $ 0$ & $ \frac{  1}{   6}$ & $ \frac{  1}{   3}$ & $ 0$ & $ 0$ & $ 0$ & $ 0$ & $ 0$ & $ 0$  \\
 7  7 & $-\frac{  1}{   9}$ & $ 0$ & $ 0$ & $ \frac{  1}{   6}$ & $ 0$ & $-\frac{  2}{   9}$ & $ 0$ & $ 0$ & $ \frac{  1}{   6}$ & $ \frac{  1}{   3}$ & $ 0$ & $ 0$ & $ 0$ & $ 0$ & $ 0$ & $ 0$  \\
\hline
\end{tabular}

\begin{tabular}{c|cccccccccccccccc}  \hline
\multicolumn{ 5}{c}{$Y= 0$~~  $I=0          $~~  $J=\frac{5}{2}$~~
$Y_4=\frac{1}{3} $}  \\  \hline
 B  M & 12214 & 22214 & 23314 & 24414 \\  \hline
 6  6 & $ \frac{  1}{   3}$ & $ \frac{  2}{   3}$ & $ 0$ & $ 0$  \\
13 18 & $ \frac{  1}{   4}$ & $-\frac{  1}{   8}$ & $ \frac{  1}{   4}$ & $-\frac{  3}{   8}$  \\
14 19 & $ \frac{  3}{   8}$ & $-\frac{  3}{  16}$ & $-\frac{  3}{   8}$ & $ \frac{  1}{  16}$  \\
15 16 & $ \frac{  1}{  24}$ & $-\frac{  1}{  48}$ & $ \frac{  3}{
8}$ & $ \frac{  9}{  16}$  \\ \hline
\end{tabular}

\begin{tabular}{c|cccccccccccccccc}  \hline
\multicolumn{ 4}{c}{$Y= 0$~~  $I=0          $~~  $J=\frac{5}{2}$~~
$Y_4=-\frac{2}{3}$}  \\  \hline
 B  M & 12215 & 22215 & 24415 \\  \hline
 7  7 & $-\frac{  1}{   3}$ & $-\frac{  2}{   3}$ & $ 0$  \\
14 20 & $-\frac{  1}{   2}$ & $ \frac{  1}{   4}$ & $ \frac{  1}{   4}$  \\
16 17 & $-\frac{  1}{   6}$ & $ \frac{  1}{  12}$ & $-\frac{  3}{
4}$  \\ \hline
\end{tabular}

\begin{tabular}{c|cccccccccccccccc}  \hline
\multicolumn{17}{c}{$Y= 0$~~  $I=0          $~~  $J=\frac{3}{2}$~~
$Y_4=\frac{1}{3} $}  \\  \hline
 B  M & 12224 & 12324 & 12424 & 21224 & 22224 & 22324 & 22424 & 12214 & 22214 & 23224 & 23424 & 24324 & 24424 & 23314 & 24224 & 24414 \\  \hline
 1  8 & $-\frac{  1}{  24}$ & $ \frac{  1}{  12}$ & $ \frac{  1}{   8}$ & $-\frac{  1}{   4}$ & $-\frac{  1}{  12}$ & $ \frac{  1}{   6}$ & $ \frac{  1}{   4}$ & $ 0$ & $ 0$ & $ 0$ & $ 0$ & $ 0$ & $ 0$ & $ 0$ & $ 0$ & $ 0$  \\
 2  9 & $-\frac{  1}{  16}$ & $-\frac{  1}{   8}$ & $-\frac{  1}{  48}$ & $-\frac{  3}{   8}$ & $-\frac{  1}{   8}$ & $-\frac{  1}{   4}$ & $-\frac{  1}{  24}$ & $ 0$ & $ 0$ & $ 0$ & $ 0$ & $ 0$ & $ 0$ & $ 0$ & $ 0$ & $ 0$  \\
 3  6 & $-\frac{  1}{ 144}$ & $ \frac{  1}{   8}$ & $-\frac{  3}{  16}$ & $-\frac{  1}{  24}$ & $-\frac{  1}{  72}$ & $ \frac{  1}{   4}$ & $-\frac{  3}{   8}$ & $ 0$ & $ 0$ & $ 0$ & $ 0$ & $ 0$ & $ 0$ & $ 0$ & $ 0$ & $ 0$  \\
 6  1 & $ \frac{  1}{  12}$ & $ 0$ & $ 0$ & $-\frac{  1}{   8}$ & $ \frac{  1}{   6}$ & $ 0$ & $ 0$ & $ \frac{  5}{  24}$ & $ \frac{  5}{  12}$ & $ 0$ & $ 0$ & $ 0$ & $ 0$ & $ 0$ & $ 0$ & $ 0$  \\
 6  6 & $-\frac{  5}{  36}$ & $ 0$ & $ 0$ & $ \frac{  5}{  24}$ & $-\frac{  5}{  18}$ & $ 0$ & $ 0$ & $ \frac{  1}{   8}$ & $ \frac{  1}{   4}$ & $ 0$ & $ 0$ & $ 0$ & $ 0$ & $ 0$ & $ 0$ & $ 0$  \\
 9 18 & $ \frac{  1}{   6}$ & $ \frac{  1}{  12}$ & $ 0$ & $ 0$ & $-\frac{  1}{  12}$ & $-\frac{  1}{  24}$ & $ 0$ & $ 0$ & $ 0$ & $ \frac{  1}{   8}$ & $-\frac{  1}{   8}$ & $ \frac{  1}{   8}$ & $-\frac{  1}{   4}$ & $ 0$ & $ 0$ & $ 0$  \\
\hline
\end{tabular}

\begin{tabular}{c|cccccccccccccccc}  \hline
\multicolumn{13}{c}{$Y= 0$~~  $I=0          $~~  $J=\frac{3}{2}$~~
$Y_4=-\frac{2}{3}$}  \\  \hline
 B  M & 12225 & 12425 & 21225 & 22225 & 22425 & 12215 & 22215 & 23225 & 23425 & 24425 & 24225 & 24415 \\  \hline
 2 10 & $ \frac{  1}{  12}$ & $-\frac{  1}{  12}$ & $ \frac{  1}{   2}$ & $ \frac{  1}{   6}$ & $-\frac{  1}{   6}$ & $ 0$ & $ 0$ & $ 0$ & $ 0$ & $ 0$ & $ 0$ & $ 0$  \\
 4  7 & $ \frac{  1}{  36}$ & $ \frac{  1}{   4}$ & $ \frac{  1}{   6}$ & $ \frac{  1}{  18}$ & $ \frac{  1}{   2}$ & $ 0$ & $ 0$ & $ 0$ & $ 0$ & $ 0$ & $ 0$ & $ 0$  \\
 7  2 & $-\frac{  1}{  12}$ & $ 0$ & $ \frac{  1}{   8}$ & $-\frac{  1}{   6}$ & $ 0$ & $-\frac{  5}{  24}$ & $-\frac{  5}{  12}$ & $ 0$ & $ 0$ & $ 0$ & $ 0$ & $ 0$  \\
 7  7 & $ \frac{  5}{  36}$ & $ 0$ & $-\frac{  5}{  24}$ & $ \frac{  5}{  18}$ & $ 0$ & $-\frac{  1}{   8}$ & $-\frac{  1}{   4}$ & $ 0$ & $ 0$ & $ 0$ & $ 0$ & $ 0$  \\
10 20 & $-\frac{  1}{   3}$ & $ 0$ & $ 0$ & $ \frac{  1}{   6}$ & $ 0$ & $ 0$ & $ 0$ & $-\frac{  1}{   4}$ & $ \frac{  1}{  12}$ & $ \frac{  1}{   6}$ & $ 0$ & $ 0$  \\
12 17 & $-\frac{  1}{   9}$ & $ 0$ & $ 0$ & $ \frac{  1}{  18}$ & $ 0$ & $ 0$ & $ 0$ & $-\frac{  1}{  12}$ & $-\frac{  1}{   4}$ & $-\frac{  1}{   2}$ & $ 0$ & $ 0$  \\
\hline
\end{tabular}

\begin{tabular}{c|cccccccccccccccc}  \hline
\multicolumn{17}{c}{$Y= 0$~~  $I=0          $~~  $J=\frac{1}{2}$~~
$Y_4=\frac{1}{3} $}  \\  \hline
 B  M & 12224 & 12234 & 12324 & 12424 & 12434 & 21224 & 21334 & 22224 & 22234 & 22324 & 22424 & 22434 & 23224 & 23334 & 23424 & 24234 \\  \hline
 1  3 & $-\frac{  1}{  32}$ & $ \frac{  1}{  32}$ & $ \frac{  1}{  16}$ & $ \frac{  3}{  32}$ & $ \frac{  1}{  32}$ & $-\frac{  3}{  16}$ & $-\frac{  1}{  16}$ & $-\frac{  1}{  16}$ & $ \frac{  1}{  16}$ & $ \frac{  1}{   8}$ & $ \frac{  3}{  16}$ & $ \frac{  1}{  16}$ & $ 0$ & $ 0$ & $ 0$ & $ 0$  \\
 1  8 & $-\frac{  1}{  96}$ & $-\frac{  3}{  32}$ & $ \frac{  1}{  48}$ & $ \frac{  1}{  32}$ & $-\frac{  3}{  32}$ & $-\frac{  1}{  16}$ & $ \frac{  3}{  16}$ & $-\frac{  1}{  48}$ & $-\frac{  3}{  16}$ & $ \frac{  1}{  24}$ & $ \frac{  1}{  16}$ & $-\frac{  3}{  16}$ & $ 0$ & $ 0$ & $ 0$ & $ 0$  \\
 2  5 & $-\frac{  3}{  64}$ & $ \frac{  3}{  64}$ & $-\frac{  3}{  32}$ & $-\frac{  1}{  64}$ & $-\frac{  1}{ 192}$ & $-\frac{  9}{  32}$ & $ \frac{  3}{  32}$ & $-\frac{  3}{  32}$ & $ \frac{  3}{  32}$ & $-\frac{  3}{  16}$ & $-\frac{  1}{  32}$ & $-\frac{  1}{  96}$ & $ 0$ & $ 0$ & $ 0$ & $ 0$  \\
 2  9 & $-\frac{  1}{  64}$ & $-\frac{  9}{  64}$ & $-\frac{  1}{  32}$ & $-\frac{  1}{ 192}$ & $ \frac{  1}{  64}$ & $-\frac{  3}{  32}$ & $-\frac{  9}{  32}$ & $-\frac{  1}{  32}$ & $-\frac{  9}{  32}$ & $-\frac{  1}{  16}$ & $-\frac{  1}{  96}$ & $ \frac{  1}{  32}$ & $ 0$ & $ 0$ & $ 0$ & $ 0$  \\
 3  1 & $-\frac{  1}{ 192}$ & $ \frac{  1}{ 192}$ & $ \frac{  3}{  32}$ & $-\frac{  9}{  64}$ & $-\frac{  3}{  64}$ & $-\frac{  1}{  32}$ & $-\frac{  3}{  32}$ & $-\frac{  1}{  96}$ & $ \frac{  1}{  96}$ & $ \frac{  3}{  16}$ & $-\frac{  9}{  32}$ & $-\frac{  3}{  32}$ & $ 0$ & $ 0$ & $ 0$ & $ 0$  \\
\hline
\end{tabular}

\begin{tabular}{c|cccccccccccccccc}  \hline
 B  M & 24324 & 24424 & 24434 & 24224 \\  \hline
 1  3 & $ 0$ & $ 0$ & $ 0$ & $ 0$  \\
 1  8 & $ 0$ & $ 0$ & $ 0$ & $ 0$  \\
 2  5 & $ 0$ & $ 0$ & $ 0$ & $ 0$  \\
 2  9 & $ 0$ & $ 0$ & $ 0$ & $ 0$  \\
 3  1 & $ 0$ & $ 0$ & $ 0$ & $ 0$  \\
 3  6 & $ 0$ & $ 0$ & $ 0$ & $ 0$  \\
 6  6 & $ 0$ & $ 0$ & $ 0$ & $ 0$  \\
 9 13 & $ \frac{  3}{  32}$ & $-\frac{  3}{  16}$ & $-\frac{  3}{  64}$ & $ 0$  \\
 9 18 & $ \frac{  1}{  32}$ & $-\frac{  1}{  16}$ & $ \frac{  9}{  64}$ & $ 0$  \\
10 15 & $-\frac{  9}{  64}$ & $ \frac{  1}{  32}$ & $ \frac{  1}{ 128}$ & $ 0$  \\
10 19 & $-\frac{  3}{  64}$ & $ \frac{  1}{  96}$ & $-\frac{  3}{ 128}$ & $ 0$  \\
11 11 & $ \frac{  9}{  64}$ & $ \frac{  9}{  32}$ & $ \frac{  9}{ 128}$ & $ 0$  \\
11 16 & $ \frac{  3}{  64}$ & $ \frac{  3}{  32}$ & $-\frac{ 27}{ 128}$ & $ 0$  \\
13 18 & $ \frac{  1}{   8}$ & $ \frac{  1}{  16}$ & $ 0$ & $-\frac{  3}{  16}$  \\
14 19 & $-\frac{  3}{  16}$ & $-\frac{  1}{  96}$ & $ 0$ & $-\frac{  9}{  32}$  \\
15 16 & $ \frac{  3}{  16}$ & $-\frac{  3}{  32}$ & $ 0$ & $-\frac{  1}{  32}$  \\
18 11 & $ 0$ & $ 0$ & $ 0$ & $ \frac{  3}{ 128}$  \\
18 16 & $ 0$ & $ 0$ & $ 0$ & $-\frac{  1}{   8}$  \\
21 15 & $ 0$ & $-\frac{  1}{   8}$ & $ \frac{  1}{   8}$ & $ 0$  \\
21 19 & $ 0$ & $-\frac{  1}{  24}$ & $-\frac{  3}{   8}$ & $ 0$
\\ \hline
\end{tabular}

\begin{tabular}{c|cccccccccccccccc}  \hline
\multicolumn{16}{c}{$Y= 0$~~  $I=0          $~~  $J=\frac{1}{2}$~~
$Y_4=-\frac{2}{3}$}  \\  \hline
 B  M & 12225 & 12235 & 12425 & 12435 & 21225 & 22225 & 22235 & 22425 & 22435 & 23225 & 23425 & 24235 & 24425 & 24435 & 24225 \\  \hline
 2  4 & $ \frac{  1}{  16}$ & $-\frac{  1}{  16}$ & $-\frac{  1}{  16}$ & $-\frac{  1}{  48}$ & $ \frac{  3}{   8}$ & $ \frac{  1}{   8}$ & $-\frac{  1}{   8}$ & $-\frac{  1}{   8}$ & $-\frac{  1}{  24}$ & $ 0$ & $ 0$ & $ 0$ & $ 0$ & $ 0$ & $ 0$  \\
 2 10 & $ \frac{  1}{  48}$ & $ \frac{  3}{  16}$ & $-\frac{  1}{  48}$ & $ \frac{  1}{  16}$ & $ \frac{  1}{   8}$ & $ \frac{  1}{  24}$ & $ \frac{  3}{   8}$ & $-\frac{  1}{  24}$ & $ \frac{  1}{   8}$ & $ 0$ & $ 0$ & $ 0$ & $ 0$ & $ 0$ & $ 0$  \\
 4  2 & $ \frac{  1}{  48}$ & $-\frac{  1}{  48}$ & $ \frac{  3}{  16}$ & $ \frac{  1}{  16}$ & $ \frac{  1}{   8}$ & $ \frac{  1}{  24}$ & $-\frac{  1}{  24}$ & $ \frac{  3}{   8}$ & $ \frac{  1}{   8}$ & $ 0$ & $ 0$ & $ 0$ & $ 0$ & $ 0$ & $ 0$  \\
 4  7 & $ \frac{  1}{ 144}$ & $ \frac{  1}{  16}$ & $ \frac{  1}{  16}$ & $-\frac{  3}{  16}$ & $ \frac{  1}{  24}$ & $ \frac{  1}{  72}$ & $ \frac{  1}{   8}$ & $ \frac{  1}{   8}$ & $-\frac{  3}{   8}$ & $ 0$ & $ 0$ & $ 0$ & $ 0$ & $ 0$ & $ 0$  \\
 7  7 & $ \frac{  2}{   9}$ & $ 0$ & $ 0$ & $ 0$ & $-\frac{  1}{   3}$ & $ \frac{  4}{   9}$ & $ 0$ & $ 0$ & $ 0$ & $ 0$ & $ 0$ & $ 0$ & $ 0$ & $ 0$ & $ 0$  \\
10 14 & $-\frac{  1}{   4}$ & $-\frac{  1}{  16}$ & $ 0$ & $ \frac{  1}{  48}$ & $ 0$ & $ \frac{  1}{   8}$ & $ \frac{  1}{  32}$ & $ 0$ & $-\frac{  1}{  96}$ & $-\frac{  3}{  16}$ & $ \frac{  1}{  16}$ & $-\frac{  3}{  32}$ & $ \frac{  1}{   8}$ & $ \frac{  1}{  32}$ & $ 0$  \\
\hline
\end{tabular}

\begin{tabular}{c|cccccccccccccccc}  \hline
\multicolumn{ 5}{c}{$Y=-1$~~  $I=\frac{3}{2}$~~  $J=\frac{5}{2}$~~
$Y_4=-\frac{2}{3}$}  \\  \hline
 B  M & 12213 & 21113 & 22213 & 23313 \\  \hline
 6  8 & $ \frac{  1}{   6}$ & $-\frac{  1}{   2}$ & $ \frac{  1}{   3}$ & $ 0$  \\
 7  6 & $-\frac{  1}{   6}$ & $-\frac{  1}{   2}$ & $-\frac{  1}{   3}$ & $ 0$  \\
15 18 & $ \frac{  1}{   3}$ & $ 0$ & $-\frac{  1}{   6}$ & $-\frac{  1}{   2}$  \\
16 16 & $ \frac{  1}{   3}$ & $ 0$ & $-\frac{  1}{   6}$ & $
\frac{  1}{   2}$  \\ \hline
\end{tabular}

\begin{tabular}{c|cccccccccccccccc}  \hline
\multicolumn{15}{c}{$Y=-1$~~  $I=\frac{3}{2}$~~  $J=\frac{3}{2}$~~
$Y_4=-\frac{2}{3}$}  \\  \hline
 B  M & 12223 & 12323 & 21223 & 22223 & 22323 & 12123 & 12213 & 21113 & 22123 & 22213 & 23223 & 24323 & 23313 & 24223 \\  \hline
 3  8 & $-\frac{  1}{  18}$ & $-\frac{  1}{   6}$ & $-\frac{  1}{   3}$ & $-\frac{  1}{   9}$ & $-\frac{  1}{   3}$ & $ 0$ & $ 0$ & $ 0$ & $ 0$ & $ 0$ & $ 0$ & $ 0$ & $ 0$ & $ 0$  \\
 4  6 & $-\frac{  1}{  18}$ & $ \frac{  1}{   6}$ & $-\frac{  1}{   3}$ & $-\frac{  1}{   9}$ & $ \frac{  1}{   3}$ & $ 0$ & $ 0$ & $ 0$ & $ 0$ & $ 0$ & $ 0$ & $ 0$ & $ 0$ & $ 0$  \\
 6  3 & $ \frac{  1}{  24}$ & $ 0$ & $-\frac{  1}{  16}$ & $ \frac{  1}{  12}$ & $ 0$ & $ \frac{  1}{  16}$ & $ \frac{  5}{  48}$ & $-\frac{  5}{  16}$ & $ \frac{  1}{   8}$ & $ \frac{  5}{  24}$ & $ 0$ & $ 0$ & $ 0$ & $ 0$  \\
 6  8 & $-\frac{  5}{  72}$ & $ 0$ & $ \frac{  5}{  48}$ & $-\frac{  5}{  36}$ & $ 0$ & $-\frac{  5}{  48}$ & $ \frac{  1}{  16}$ & $-\frac{  3}{  16}$ & $-\frac{  5}{  24}$ & $ \frac{  1}{   8}$ & $ 0$ & $ 0$ & $ 0$ & $ 0$  \\
 7  1 & $-\frac{  1}{  24}$ & $ 0$ & $ \frac{  1}{  16}$ & $-\frac{  1}{  12}$ & $ 0$ & $ \frac{  1}{  16}$ & $-\frac{  5}{  48}$ & $-\frac{  5}{  16}$ & $ \frac{  1}{   8}$ & $-\frac{  5}{  24}$ & $ 0$ & $ 0$ & $ 0$ & $ 0$  \\
 7  6 & $ \frac{  5}{  72}$ & $ 0$ & $-\frac{  5}{  48}$ & $ \frac{  5}{  36}$ & $ 0$ & $-\frac{  5}{  48}$ & $-\frac{  1}{  16}$ & $-\frac{  3}{  16}$ & $-\frac{  5}{  24}$ & $-\frac{  1}{   8}$ & $ 0$ & $ 0$ & $ 0$ & $ 0$  \\
\hline
\end{tabular}

\begin{tabular}{c|cccccccccccccccc}  \hline
\multicolumn{17}{c}{$Y=-1$~~  $I=\frac{3}{2}$~~  $J=\frac{1}{2}$~~
$Y_4=-\frac{2}{3}$}  \\  \hline
 B  M & 12223 & 12233 & 12323 & 21223 & 21333 & 22223 & 22233 & 22323 & 12123 & 22123 & 23223 & 23333 & 24233 & 24323 & 24223 & 23133 \\  \hline
 3  3 & $-\frac{  1}{  24}$ & $ \frac{  1}{  24}$ & $-\frac{  1}{   8}$ & $-\frac{  1}{   4}$ & $ \frac{  1}{   8}$ & $-\frac{  1}{  12}$ & $ \frac{  1}{  12}$ & $-\frac{  1}{   4}$ & $ 0$ & $ 0$ & $ 0$ & $ 0$ & $ 0$ & $ 0$ & $ 0$ & $ 0$  \\
 3  8 & $-\frac{  1}{  72}$ & $-\frac{  1}{   8}$ & $-\frac{  1}{  24}$ & $-\frac{  1}{  12}$ & $-\frac{  3}{   8}$ & $-\frac{  1}{  36}$ & $-\frac{  1}{   4}$ & $-\frac{  1}{  12}$ & $ 0$ & $ 0$ & $ 0$ & $ 0$ & $ 0$ & $ 0$ & $ 0$ & $ 0$  \\
 4  1 & $-\frac{  1}{  24}$ & $ \frac{  1}{  24}$ & $ \frac{  1}{   8}$ & $-\frac{  1}{   4}$ & $-\frac{  1}{   8}$ & $-\frac{  1}{  12}$ & $ \frac{  1}{  12}$ & $ \frac{  1}{   4}$ & $ 0$ & $ 0$ & $ 0$ & $ 0$ & $ 0$ & $ 0$ & $ 0$ & $ 0$  \\
 4  6 & $-\frac{  1}{  72}$ & $-\frac{  1}{   8}$ & $ \frac{  1}{  24}$ & $-\frac{  1}{  12}$ & $ \frac{  3}{   8}$ & $-\frac{  1}{  36}$ & $-\frac{  1}{   4}$ & $ \frac{  1}{  12}$ & $ 0$ & $ 0$ & $ 0$ & $ 0$ & $ 0$ & $ 0$ & $ 0$ & $ 0$  \\
 6  8 & $-\frac{  1}{   9}$ & $ 0$ & $ 0$ & $ \frac{  1}{   6}$ & $ 0$ & $-\frac{  2}{   9}$ & $ 0$ & $ 0$ & $-\frac{  1}{   6}$ & $-\frac{  1}{   3}$ & $ 0$ & $ 0$ & $ 0$ & $ 0$ & $ 0$ & $ 0$  \\
 7  6 & $ \frac{  1}{   9}$ & $ 0$ & $ 0$ & $-\frac{  1}{   6}$ & $ 0$ & $ \frac{  2}{   9}$ & $ 0$ & $ 0$ & $-\frac{  1}{   6}$ & $-\frac{  1}{   3}$ & $ 0$ & $ 0$ & $ 0$ & $ 0$ & $ 0$ & $ 0$  \\
\hline
\end{tabular}%
\end{tiny}

\vspace{8mm}

Table A2. Indices of baryons and mesons. The subscript 8 means color
octet; the superscript $\frac{1}{2}$ refers to the spin of baryon.
$\Lambda_{s}$ is the flavor singlet $\Lambda$.

%

\begin{references}
\bibitem{qcw} D. Qing, X.S. Chen and F. Wang, Phys. Rev. {\bf C57}, R31 (1998);
{\bf D58}, 114032 (1998).
\bibitem{BSZou} B.S. Zou and D.O. Riska, Phys. Rev. Lett. {\bf
 95}, 072001 (2005).
\bibitem{tetraquark}B. Aubert, {\em et al}, Phys. Rev. Lett. {\bf 90}, 242001 (2003);
 S.K. Choi, {\em et al}, Phys. Rev. Lett. {\bf 91}, 262001 (2003);
 A.V. Evdokimov, {\em et al} (SELEX Collaboration) Phys. Rev. Lett.
{\bf 93}, 242001 (2004); B. Aubert, {\em et al}, Phys. Rev. Lett.
{\bf 95}, 142001 (2005); R.L. Jaffe, Phys. Report {\bf 409},1
(2005).
\bibitem{dibaryon}
 F. Wang, J.L. Ping and T. Goldman, Phys. Rev. {\bf C51}, 3411 (1995);
 T. Goldman, K. Maltman, G.J. Stephenson, Jr, J.L.Ping and F. Wang,
 Mod. Phys. Lett. {\bf A13}, 59 (1998);
 J.L. Ping, F. Wang and T. Goldman, Nucl. Phys. {\bf A688} 871 (2001);
 H.R. Pang, J.L. Ping, F. Wang and T. Goldman, Phys. Rev. {\bf C65},
 014003 (2001).
 J.L. Ping, H.R. Pang, F. Wang and T. Goldman, Phys. Rev. {\bf C65}, 044003 (2002).
\bibitem{group1} H.X. Huang, C.R. Deng, J.L. Ping, F. Wang and T. Goldman, Phys. Rev. {\bf
C77}, 025201 (2008).
\bibitem{Isgur} A. De Rujula, H. Georgi and S.L. Glashow, Phys. Rev. {\bf D12}, 147 (1975);
 N. Isgur and G. Karl, Phys. Rev. {\bf D18}, 4178 (1978); {\bf D19}, 2653
 (1979); {\bf D20}, 1191 (1979).
\bibitem{chiral}A. Valcarce, H. Garcilazo, F. Fernandez, and P. Gonzalez,
Rept. Prog. Phys. {\bf 68} (2005) 965.
\bibitem{qdcsm} F. Wang, G.H. Wu and L.J. Teng, {\it Phys. Rev. Lett. }{\bf 69}, 2901 (1992);
G.H. Wu, J.L. Ping, L.J. Teng, F. Wang and T. Goldman, Nucl. Phys.
{\bf A673}, 279 (2000); L.Z. Chen, H.R. Pang, H.X. Huang, J.L. Ping
and F. Wang, Phys. Rev. {\bf C76}, 014001 (2007).
\bibitem{ISFBook} J. Q. Chen {\em et al.}, {\it Tables of the
Clebsch-Gordan, Racah and Subduction Coefficients of $\mbox{SU}(n)$
Groups} (World Scientific, Singapore, 1987);  {\it Tables of the
$\mbox{SU}(mn)\supset \mbox{SU}(m)\times \mbox{SU}(n)$ Coefficients
of Fractional Parentage} (World Scientific, Singapore, 1991).
\bibitem{Harvey} M. Harvey,  Nucl. Phys. {\bf A352}, 301 (1981); {\bf 481} 834 (1988).
\bibitem{chen1}J.Q. Chen, Y.J. Shi, D.H. Feng and M. Vallieres, Nucl. Phys. {\bf
A393}, 122 (1983).
\bibitem{qdcsm1}F. Wang, J. L. Ping, and T. Goldman, {\it Phys. Rev. C}{\bf 51}, 1648 (1995).
\bibitem{JW}R. Jaffe and F. Wilczek, Phys. Rev. Lett. {\bf 91}, 232003(2003).
\bibitem{Zhang} F. Huang, Z. Y. Zhang and Y. W. Yu {\it Phys. Rev. }{\bf C 70}, 044004 (2004).
\bibitem{Zhang2} F. Huang, Z. Y. Zhang and Y. W. Yu {\it Phys. Rev. }{\bf C 73}, 025207
(2006); \\
F. Huang and Z. Y. Zhang {\it Phys. Rev. }{\bf C 72}, 068201
(2005).
\bibitem{meson} J. Vijande, F Fern\'{a}ndez and A. Valcarce,
{\it J. Phys. }{\bf G31}, 481 (2005).
\end{references}
\end{document}